\documentclass[a4paper,12pt]{article}

\usepackage{amsmath}
\usepackage{amsthm}
\usepackage{setspace}
\usepackage{graphicx}
\usepackage{authblk}
\usepackage{enumitem}
\usepackage{comment}
\usepackage{amsfonts}
\usepackage{natbib}
\usepackage{float}
\usepackage{appendix}
\usepackage{hyperref}
\graphicspath{
{./figure/}
}

\usepackage[showframe=false]{geometry}
\usepackage{changepage}
\usepackage[caption=false]{subfig}%

\newtheorem{theorem}{Theorem}
\newcommand{\vect}[1]{\ensuremath{\boldsymbol{#1}}}
\newcommand{\mat}[1]{\ensuremath{\boldsymbol{#1}}}

\begin{document}
\title{Robust Estimation under Linear Mixed Models: The Minimum Density Power Divergence Approach}
\date{}
\author[1]{Giovanni Saraceno}
\author[2]{Abhik Ghosh}
\author[2]{Ayanendranath Basu}
\author[1]{Claudio Agostinelli}
\affil[1]{Department of Mathematics, University of Trento, Trento, Italy}
\affil[2]{Interdisciplinary Statistical Research Unit, Indian Statistical Institute, Kolkata, India}
\maketitle

\begin{abstract}
Many real-life data sets can be analyzed using Linear Mixed Models (LMMs). Since these are ordinarily based on normality assumptions, under small deviations from the model the inference can be highly unstable when the associated parameters are estimated by classical methods. On the other hand, the density power divergence (DPD) family, which measures the discrepancy between two probability density functions, has been successfully used to build robust estimators with high stability associated with minimal loss in efficiency. Here, we develop the minimum DPD estimator (MDPDE) for independent but non identically distributed observations in LMMs. We prove the theoretical properties, including consistency and asymptotic normality. The influence function and sensitivity measures are studied to explore the robustness properties. As a data based choice of the MDPDE tuning parameter $\alpha$ is very important, we propose two candidates as “optimal” choices, where optimality is in the sense of choosing the strongest downweighting that is necessary for the particular data set. We conduct a simulation study comparing the proposed MDPDE, for different values of $\alpha$, with the S-estimators, M-estimators and the classical maximum likelihood estimator, considering different levels of contamination. Finally, we illustrate the performance of our proposal on a real-data example.

\bigskip\noindent \textbf{Keywords:} Linear Mixed Models, Minimum Density Power Divergence Estimator, Robustness.

%
\end{abstract}

\newpage
\section{Introduction}
\label{sec:introduction}

A major interest in statistics concerns the estimation of averages and their variation. 
The most commonly used method for this purpose is, probably,  the \textit{Linear Model} (LM).
In this model, to give an example from a two way layout, the expected value (mean) $\mu_{ij}$ of an observation $y_{ij}$, 
may be expressed as a linear combination of unknown parameters such as $\mu_{ij} = \mu + \alpha_i + \beta_j$, 
where $\mu$, $\alpha_i$ and $\beta_j$ are the constants which we are interested in estimating. 
The linearity in the parameters means that we can write a linear model in the form $\vect{y}_i = \mat{X}_i\vect{\beta} + \vect{\epsilon_i}$, 
where $\vect{\beta}$ is the vector of unknown parameters and the $\mat{X}_i$s are known matrices. 
This formulation is the same as that used in case of linear regression model. 
In the present work, we consider the \textit{Linear Mixed Models} (LMMs), 
in which some (unknown) parameters are not treated as constants but as random variables. 
Random terms come into play when some items cannot be considered as fixed quantities, although their distributions are of interest. 
Hence, they are the tools to generalize the results to the entire population under study. 
The types of data that may be appropriately analyzed by LMMs include (i) \emph{Clustered data} where the dependent variable 
is measured once for each subject (the unit of analysis) and the units of analysis are grouped into, or nested within, clusters; 
(ii) \emph{Repeated-measures data} where the dependent variable is measured more than once on the same unit of analysis across levels of a factor, 
which may be time or experimental conditions; (iii) \emph{Longitudinal data} where the dependent variable is measured at several points 
in time for each unit of analysis. For a general review of LMs and LMMs see \citet{McCulloch2001}.

The standard methods used to estimate the parameters in LMMs are methods of  maximum likelihood and restricted maximum likelihood. 
Generally, LMMs are based on normality assumptions and it is well known that these classical methods are not robust 
and can be greatly affected in the presence of small deviations from the assumptions. 
Furthermore, outlier detection for modern large data sets can be very challenging and, 
in any case, robust techniques cannot be replaced by the application of classical methods on outlier deleted data.

To answer the need for robust estimation in linear mixed models, a few methods have been proposed. 
The initial attempts were based on weighted versions of the log-likelihood function (see \citet{Huggins1993a}, \citet{Huggins1993b}, \citet{Huggins1994}, \citet{Stahel1997}, \citet{Richardson1995}, \citet{Richardson1997}, \citet{Welsh1997}). 
Another attempt, discussed in \citet{Welsh1997}, of robustifying linear mixed models consists of replacing the Gaussian distribution 
by the Student's \textit{t} distribution (see also \citet{Lange1989}, \citet{Pinheiro2001}). However, this modification of the error distribution is intractable and complicated to implement. 
\citet{Copt2006} adapted a multivariate high breakdown point S-estimator, namely CVFS-estimator, to the linear mixed models setup, 
while the estimator given by \citet{Koller2013}, namely SMDM-estimator, attempts to achieve robustness by a robustification of the score equations. Robust estimators have been proposed, more generally, for generalized linear mixed models by \citet{Yau2002} and \citet{Sinha2004}. 

The density power divergence (DPD) \citep{Basu1998}, 
which measures the discrepancy between two probability density functions, has been successfully used to build a robust estimator 
for independent and identically distributed observations. 
\cite{Ghosh2013} extended the construction of the DPD and the corresponding minimum DPD estimator (MDPDE) 
to the case of independent but non-identically distributed data. 
This approach and theory covers the linear regression model, and has later been extended to more general parametric regression models
(\citet{Ghosh2016}, \citet{Ghosh2019b}; \citet{Castilla2018}, \citet{Castilla2019}; \citet{Ghosh2019a}, etc.).
This MDPDE has become widely popular in recent times due to its good (asymptotic) efficiency along with high robustness,
easy computability and direct interpretation as an intuitive generalization of the maximum likelihood estimator (MLE).

In the present work, we aim to develop a general robust estimation procedure that is able to deal with the linear mixed model setup. 
Hence, we adapt the MDPDE in order to treat LMMs where the data are independent but non-identically distributed. 
We prove that the introduced estimator satisfies the robustness properties as well as 
the recommended asymptotic properties of an estimator at the model.

The rest of the paper is organized as follows. In Section \ref{sec:mdpde} we briefly present the MDPDE for non-homogeneous observations. 
In Section \ref{sec:LMM} we define the proposed estimator in case of linear mixed models 
and its asymptotic and robustness properties are exploited. Section \ref{sec:montecarlo} reports the simulation study we conducted, 
comparing the performance of the MDPDE to the most recent methods, exploring also the case of contaminated data. 
Section \ref{sec:example} provides the application of the proposed estimator to a real data example. 
Concluding remarks are presented in Section \ref{sec:conclusions}. 
For brevity,  the assumptions needed to prove the asymptotic normality of the estimator, the proof of the main theorem
and some additional theoretical and Monte-Carlo results are presented in the Online Supplementary Material.

\section{The MDPDE for independent non-homogeneous observations}
\label{sec:mdpde}

The density power divergence family was first introduced by \citet{Basu1998} as a measure of discrepancy between two probability density functions.
The authors used this measure to robustly estimate the model parameters under the usual setup of independent and identically distributed data. 
The density power divergence measure $d_\alpha(g,f)$ between two probability densities $g$ and $f$ is defined, 
in terms of a single tuning parameter $\alpha \ge 0$, as
\begin{align}
  & d_\alpha(g,f) = \int \left\{ f^{1+\alpha} - \left(1 + \frac{1}{\alpha}\right)f^\alpha g + \frac{1}{\alpha}g^{1+\alpha}\right\} 
  & \mbox{ if } \alpha > 0, \label{eqn:dpd}\\
  & d_0 (g,f) = \int g \ln \left(\frac{g}{f}\right) & \mbox{ if } \alpha = 0,
\label{eqn:dpd0}
\end{align}
where $\ln$ denotes the natural logarithm. 
\citet{Basu1998} demonstrated that the tuning parameter $\alpha$ controls the trade-off between efficiency and robustness of the resulting estimator. 
With increasing $\alpha$, the estimator acquires greater stability with a slight loss in efficiency. 
Since the divergence is not defined for $\alpha = 0$, 
$d_0(g,f)$ in Equation (\ref{eqn:dpd0}) represents the divergence obtained in the limit of (\ref{eqn:dpd}) as $\alpha \rightarrow 0$, 
which corresponds to a version of the Kullback-Leibler divergence. 
On the other hand, $\alpha = 1$ generates the squared $L_2$ distance.

Let $G$ be the true data generating distribution and $g$ the corresponding density function. 
To model $g$, consider the parametric family of densities $\mathcal{F}_{\vect{\theta}} = \{f_{\vect{\theta}} : \vect{\theta} \in \Theta \subseteq \mathbb{R}^p \}$. 
The minimizer of $d_\alpha(g,f_{\vect{\theta}})$ over $\vect{\theta} \in \Theta$, whenever it exists, is the minimum DPD functional at the distribution point $G$. 
Note that, the third term  of the divergence $d_\alpha(g,f_{\vect{\theta}})$ is independent of $\vect{\theta}$, 
hence it can be discarded from the objective function as it has no role in the minimization process. 
Consider a sequence of independent and identically distributed (i.i.d) observations $\vect{Y}_1, \ldots, \vect{Y}_n$ from the true distribution $G$. 
Using the empirical distribution function $G_n$ in place of $G$, the MDPDE of $\vect{\theta}$ can be obtained by minimizing
\begin{equation*}
  \int f_{\vect{\theta}}^{1+\alpha} - \left(1+\frac{1}{\alpha}\right) \frac{1}{n}\sum_{i=1}^n f_{\vect{\theta}}^\alpha(\vect{Y}_i)
\end{equation*}
over $\vect{\theta} \in \Theta$. In the above equation, the empirical distribution function is used to approximate its theoretical version 
(or, alternatively, the sample mean is used to approximate the population mean). 
Note that, it is valid in case of  continuous densities also. See \citet{Basu2011} for more details and examples.

\citet{Ghosh2013} generalized the above concept of robust minimum DPD estimation to the more general case of independent non-homogeneous observations, 
i.e., they considered the case where the observed data $\vect{Y}_1, \ldots, \vect{Y}_n$ are independent but for each $i$, 
$\vect{Y}_i \sim g_i$ where $g_1, \ldots, g_n$ are possibly different densities with respect to some common dominating measure. 
We model $g_i$ by the family $\mathcal{F}_{i,\vect{\theta}} = \{f_i(\cdot,\vect{\theta}) : \vect{\theta} \in \Theta\}$ for $i = 1, \ldots, n$. 
While the distributions $f_i(\cdot,\vect{\theta})$ can be distinct, they share the same parameter vector $\vect{\theta}$. 
\citet{Ghosh2013} proposed to minimize the average divergence between the data points and the model densities 
which leads to the minimization of the objective function
\begin{equation}
  \label{eqn:obj-function}
  H_n(\vect{\theta})=\frac{1}{n} \sum_{i=1}^n \left[\int f_i(\vect{y},\vect{\theta})^{1+\alpha}dy - \left(1+\frac{1}{\alpha}\right)f_i(\vect{Y}_i,\vect{\theta})^\alpha 
  \right] = \frac{1}{n}\sum_{i=1}^n H_i(\vect{Y}_i,\vect{\theta}),
\end{equation}
where $H_i(\vect{Y}_i,\vect{\theta})$ is the indicated term within the square brackets in the above equation. 
Differentiating the above expression with respect to $\vect{\theta}$ we get the estimating equations of the MDPDE for non-homogeneous observations. 
Note that, the estimating equation is unbiased when each $g_i$ belongs to the model family $\mathcal{F}_{i,\vect{\theta}}$, respectively. 
When $\alpha \rightarrow 0$, the corresponding objective function reduces to
$ - \sum_{i=1}^n \ln(f_i(\vect{Y}_i,\vect{\theta})) /n $, which is the negative of the log-likelihood function. 
In Section SM--1, we report the assumptions (A1)-(A7) 
which are used to prove the asymptotic normality of the MDPDE \citep{Ghosh2013}. 

\section{The MDPDE for Linear Mixed Models}
\label{sec:LMM}

The general formulation of an LMM may be expressed as  $\vect{Y} = \mat{X}\vect{\beta} + \mat{Z} \vect{u} + \vect{\epsilon}$, 
where $\mat{X}$ and $\mat{Z}$ are known design matrices, $\vect{\beta}$ is the parameter vector for fixed effects, $\vect{u}$ is the vector of random effects
and $\vect{\epsilon}$ is the random error vector. 
More explicitly, let the model have $r$ random factors $u_j$ with $q_j$ levels, $j=1,\ldots,r$,
and denote the size of the $i$-th group by $n_i$. Note that $\sum_{i=1}^nn_i = N$ is the total number of observations. 
Under this setup we can rewrite the model as
\begin{equation}
  \label{eqn:lmm}
  \vect{Y}_i = \mat{X}_i\vect{\beta} + \sum_{j=1}^r\mat{Z}_{ij}\vect{u}_{ij} + \vect{\epsilon}_i, ~~~~~~~i=1, \ldots, n,
\end{equation}
where $\vect{Y}_i$ $(n_i \times 1)$ is the response vector for group $i$, $\mat{X}_i$ $(n_i \times k)$ and $\mat{Z}_{ij}$ $(n_i \times q_j)$ are the model matrices, 
$\vect{\beta}$ $(k \times 1)$ is the vector of unknown parameters for fixed effects, 
$\vect{u}_{ij}$ represents the realized values of $u_j$ for the $i$-th group and $\vect{\epsilon}_i$ $(n_i \times 1)$ is the error term. 
We assume that $\vect{\epsilon}_i \sim N(0,\sigma_0^2\mat{I}_{n_i})$ and $\vect{u}_{ij} \sim N(0,\sigma_j^2\mat{I}_{q_j})$, 
where $\mat{I}_n$ is the $n \times n$ identity matrix, and $\vect{\epsilon}_i$ and $\vect{u}_{ij}$ are independent of each other for all $i,j$. 
Then
\begin{equation*}
  \vect{Y}_i \sim N_{n_i}\left(\mat{X}_i\vect{\beta}, \sigma_0^2\mat{I}_{n_i} + \sum_{j=1}^r\mat{Z}_{ij}\mat{Z}_{ij}^\top\sigma^2_j\right),
  ~~~~~~ i=1, \ldots, n. 
\end{equation*}
Thus, $\vect{Y}_i$s are independent but not identically distributed; for each $i$, the covariance matrix of $\vect{Y}_i$ can be rewritten as
\begin{equation*}
  \label{eqn:Vi-matrix}
  \mat{V}_i = \sigma_0^2\mat{I}_{n_i} + \sum_{j=1}^r\mat{Z}_{ij}\mat{Z}_{ij}^\top\sigma^2_j = \sigma_0^2\left(\mat{I}_{n_i} + \sum_{j=1}^r \mat{Z}_{ij}\mat{Z}_{ij}^\top \gamma_j \right), 
  ~~~~\mbox{ with }~\gamma_j = \frac{\sigma^2_j}{\sigma_0^2}.
\end{equation*}

In this setting, we can obtain the MDPDE for the parameter vector $\vect{\theta} = (\vect{\beta}^\top; \sigma_j^2 ,j=0,1, \ldots,r)^\top$ 
by minimizing the objective function given in Equation (\ref{eqn:obj-function}) with $f_i \equiv N(\mat{X}_i\vect{\beta}, \mat{V}_i)$. 
Upon simplification, the objective function is given by
\begin{equation}
  \label{eqn:obj-func-lmm}
  H_n(\vect{\theta}) = \frac{1}{n}\sum_{i=1}^n \bigg[ \frac{1}{(2\pi)^{\frac{n_i\alpha}{2}}|\mat{V}_i|^{\frac{\alpha}{2}}(\alpha+1)^{\frac{n_i}{2}}} -\left(1+\frac{1}{\alpha}\right) \frac{e^{-\frac{\alpha}{2}(\vect{Y}_i - \mat{X}_i\vect{\beta})^\top \mat{V}_i^{-1}(\vect{Y}_i - \mat{X}_i\vect{\beta})}}{(2\pi)^{\frac{n_i\alpha}{2}}|\mat{V}_i|^{\frac{\alpha}{2}}} \bigg].
\end{equation}
Differentiating the above equation with respect to $\vect{\beta}$, we get the corresponding estimating equation for the MDPDE of $\vect{\beta}$ as 
\begin{equation}
  \label{eqn:beta-derivative}
  \frac{\partial H_n}{\partial \vect{\beta}} = \frac{1}{n} \sum_{i=1}^n\bigg[-\left(1+\frac{1}{\alpha}\right)\frac{e^{-\frac{\alpha}{2}(\vect{Y}_i - \mat{X}_i\vect{\beta})^\top \mat{V}_i^{-1}(\vect{Y}_i - \mat{X}_i\vect{\beta})}}{(2\pi)^{\frac{n_i\alpha}{2}}|\mat{V}_i|^{\frac{\alpha}{2}}}\alpha \mat{X}_i^\top \mat{V}_i^{-1}(\vect{Y}_i - \mat{X}_i\vect{\beta})\bigg] = 0.
\end{equation}
Let $\mat{U}_{ij}$ denote the partial derivative of the matrix $\mat{V}_i$ with respect to $\sigma^2_j$. 
We have that $\mat{U}_{i0} = \mat{I}_{n_i}$, and $\mat{U}_{ij} = \mat{Z}_{ij}\mat{Z}_{ij}^\top$, $j=1,\ldots,r$. 
Then, the partial derivative of the objective function with respect to $\sigma^2_j$, $j=0,1, \ldots,r$, 
leads to their MDPDE estimating equations as given by
\begin{align}
  \label{eqn:sigj-derivative}
   \frac{\partial H_n}{\partial \sigma^2_j} = \frac{1}{n} \sum_{i=1}^n 
  \bigg\{- & \frac{\alpha \mathrm{Tr}(\mat{V}_i^{-1}\mat{U}_{ij})}{2(2\pi)^{\frac{n_i\alpha}{2}}|\mat{V}_i|^{\frac{\alpha}{2}}(\alpha+1)^{\frac{n_i}{2}}}   + 
  \frac{\alpha}{2}   \left(1+\frac{1}{\alpha}\right)   \frac{e^{-\frac{\alpha}{2}(\vect{Y}_i - \mat{X}_i\vect{\beta})^\top \mat{V}_i^{-1}(\vect{Y}_i - \mat{X}_i\vect{\beta})}}{(2\pi)^{\frac{n_i\alpha}{2}}|\mat{V}_i|^{\frac{\alpha}{2}}} \nonumber \\
  & \times \Big[ \mathrm{Tr}(\mat{V}_i^{-1}\mat{U}_{ij})  - (\vect{Y}_i - \mat{X}_i\vect{\beta})^\top \mat{V}_i^{-1}\mat{U}_{ij}\mat{V}_i^{-1}(\vect{Y}_i - \mat{X}_i\vect{\beta}) \Big]\bigg\} = 0,
\end{align}
where $\mathrm{Tr}(\cdot)$ denotes the trace of the argument matrix.
Solving Equations (\ref{eqn:beta-derivative})--(\ref{eqn:sigj-derivative}) numerically we can obtain the estimates of 
$\vect{\theta} = (\vect{\beta}^\top; \sigma_j^2 ,j=0,1, \ldots,r)^\top$.
In case multiple roots exist, we should chose the one minimizing the objective function $H_n(\vect{\theta})$ in (\ref{eqn:obj-func-lmm})
as the targeted MDPDE of $\vect{\theta}$. 

Note that, substituting $\alpha=0$ in the estimating equations in  (\ref{eqn:obj-func-lmm})--(\ref{eqn:sigj-derivative}), 
we get back the MLE score equations. Thus, the MDPDE at $\alpha=0$ is nothing but the usual MLE also under our LMMs.

\subsection{Asymptotic efficiency}
\label{subsec:asymtotics}

We assume that the true densities $g_i$ belong to the model family, i.e. $g_i=f_i(\cdot,\vect{\theta})$ for some value of $\vect{\theta} \in \Theta$. 
At first, note that, for each $i$, the score function for the LMM (\ref{eqn:lmm}) is given by 
\begin{equation*}
  \label{eqn:score-function}
  u_i(\vect{y}_i;\vect{\theta}) = \left ( \begin{array}{c}
    u_i^{(1)}(\vect{y}_i;\vect{\theta}) \\
    u_i^{(2)}(\vect{y}_i;\vect{\theta})
  \end{array} \right ),
\end{equation*}
where $u_i^{(1)}(\vect{y}_i;\vect{\theta}) = \mat{X}_i^\top \mat{V}_i^{-1} (\vect{Y}_i - \mat{X}_i\vect{\beta})$ 
and the $j$-th element of the $(r+1)$-vector $u_i^{(2)}(\vect{y}_i;\vect{\theta})$ is given by
$\left[\frac{\mathrm{Tr}(\mat{V}_i^{-1}\mat{U}_{ij})}{2} -\frac{1}{2}(\vect{Y}_i - \mat{X}_i\vect{\beta})^\top \mat{V}_i^{-1}\mat{U}_{ij}\mat{V}_i^{-1}(\vect{Y}_i - \mat{X}_i\vect{\beta})\right]$ for $j=0, 1, \ldots, r$. 
Now, let us fix an $\alpha\geq 0$. For each $i$, we define  
\begin{equation*}
\mat{J}^{(i)} = \left ( \begin{array}{cc}
\mat{J}^{(i)}_{11} & 0 \\
0 & \mat{J}^{(i)}_{22}
\end{array} \right ),~~~~~~
\mat{\Omega}^{(i)} = \left ( \begin{array}{ccc}
\mat{\Omega}^{(i)}_{11} & 0  \\
0 & \mat{\Omega}^{(i)}_{22}
\end{array} \right ),
~~\mbox{ and }~~
\vect{\xi}_i = \left ( \begin{array}{c}
0 \\
\vect{\xi}_i^{(2)}
\end{array} \right ),
\end{equation*}
where 
\begin{eqnarray*}
 \label{eqn:ji-matrix}
\mat{J}^{(i)}_{11} &=& \frac{4\mat{X}_i^\top \mat{V}_i^{-1}\mat{X}_i}{(1+\alpha)^{\frac{n_i}{2}+1}}, 
\\
(j,k)\mbox{-th element of  } \mat{J}^{(i)}_{22} &=& \frac{T(\mat{V}_i^{-1}\mat{U}_{ij},\mat{V}_i^{-1}\mat{U}_{ik})}{(1+\alpha)^{\frac{n_i}{2}+2}}, 
\nonumber\\
\label{eqn:omegai-matrix}
\mat{\Omega}^{(i)}_{11} &=& \frac{4\mat{X}_i^\top \mat{V}_i^{-1}\mat{X}_i}{(1+2\alpha)^{\frac{n_i}{2}+1}}, 
\\
(j,k)\mbox{-th element of  } \mat{\Omega}^{(i)}_{22} &=& \frac{T(4,\mat{V}_i^{-1}\mat{U}_{ij},\mat{V}_i^{-1}\mat{U}_{ik})}{(1+2\alpha)^{\frac{n_i}{2}+2}} -\frac{\alpha^2\mathrm{Tr}(\mat{V}_i^{-1}\mat{U}_{ik})\mathrm{Tr}(\mat{V}_i^{-1}\mat{U}_{ij})}{(1+\alpha)^{n_i+2}},
\nonumber\\
\label{eqn:xi-vector}
j\mbox{-th element of  } \vect{\xi}_i^{(2)} &=& \frac{\eta_{i\alpha}\mathrm{Tr}(\mat{V}_i^{-1}\mat{U}_{ij})\alpha}{2(1+\alpha)^{\frac{n_i}{2}+1}},
\end{eqnarray*}
for $j, k = 0,1, \ldots, r$, with $\eta_{i\alpha} =  {(2\pi)^{-\frac{n_i\alpha}{2}}|\mat{V}_i|^{-\frac{\alpha}{2}}}$ and
$T(c,\mat{A},\mat{B}) = c\alpha^2\mathrm{Tr}(\mat{A})\mathrm{Tr}(\mat{B}) +2\mathrm{Tr}(\mat{AB})$ 
for general matrices $\mat{A},\mat{B}$  and a constant $c$  ($c=1$ if not specified). 
Finally, put 
$$
\mat{\Psi}_n = \frac{1}{n} \sum_{i=1}^n \frac{\eta_{i\alpha}}{4} \mat{J}^{(i)}
~~~~\mbox{ and }~~ 
\mat{\Omega}_n = \frac{1}{n} \sum_{i=1}^n  \frac{\eta_{i\alpha}^2}{4} \mat{\Omega}^{(i)}.
$$ 

Now, we present some conditions on the independent variables and on the variance-covariance matrices that will be used to derive the asymptotic distribution of the MDPDE of the parameter vector $\vect{\theta} = (\vect{\beta}^\top; \sigma_j^2 ,j=0,1, \ldots,r)^\top$  in the LMM application.

\begin{enumerate}[label=\text{(MM\arabic*)}]
\item \label{hp:mm1} Define $\mat{X}_i^\prime = \left ( \frac{\eta_{i\alpha}\mat{V}_i^{-1}}{(1+\alpha)^{\frac{n_i}{2}+1}} \right )^{\frac{1}{2}}\mat{X}_i$, 
for each $i$,  and ${\mat{X}^\prime}=\mbox{Block-Diag}\{\mat{X}_i^\prime : i=1,\ldots, n \}$.
Then the $\mat{X}^\prime$ matrix satisfies
  \begin{equation}
    \inf_n \left [\mbox{min eigenvalue of } \frac{{\mat{X}^\prime}^\top \mat{X}^\prime}{n}\right ] > 0,
  \end{equation}
  and $\mat{X}_i$ and $\mat{Z}_i$ are full rank matrices for all $i$.
  
\item \label{hp:mm2} The values of $\mat{X}_i$'s are such that,  for all $j,k,l$
  \begin{equation}
    \label{eqn:mm2-a6-beta}
    \sup_{n>1} \max_{1 \le i\le n} |\vect{X}_{ij}^\top \mat{V}_i^{-\frac{1}{2}}| = O(1), \qquad \sup_{n>1} \max_{1 \le i\le n} |\vect{X}_{ij}^\top \mat{V}_i^{-1}\vect{X}_{ik}| = O(1),
  \end{equation}
  \begin{equation}
    \label{eqn:mm2-a5}
    \frac{1}{n}\sum_{i=1}^{n} |\vect{X}_{ij}^\top \mat{V}_i^{-1} \vect{X}_{ik}\vect{X}_{il}^\top \mat{V}_i^{-\frac{1}{2}}\textbf{1}| = O(1), 
    \quad 
    \frac{1}{n} \sum_{i=1}^n |\vect{X}_{ij}^\top \mat{V}_i^{-\frac{1}{2}}|diag(\mat{V}_i^{-\frac{1}{2}}\vect{X}_{ik}\vect{X}_{il}^\top \mat{V}_i^{-\frac{1}{2}})\textbf{1} = O(1),
  \end{equation}
  where $\vect{1}(n_i \times 1)$ is a vector of $1$'s. 
  
\item \label{hp:mm3} The matrices $\mat{V}_i$ and $\mat{U}_{ij}$ are such that, for all $j,k,l = 0, \ldots, r$,
  \begin{align}
    \label{eqn:mm3-a6}
    \frac{1}{n}\sum_{i=1}^n \mathrm{Tr}(\mat{V}_i^{-1}\mat{U}_{ij})& = O(1), 
    \qquad \frac{1}{n}\sum_{i=1}^n \mathrm{Tr}(\mat{V}_i^{-1}\mat{U}_{ij})\mathrm{Tr}(\mat{V}_i^{-1}\mat{U}_{ik}) = O(1), \\
& \frac{1}{n}\sum_{i=1}^n \mathrm{Tr}(\mat{V}_i^{-1}\mat{U}_{ik}\mat{V}_i^{-1}\mat{U}_{ij}) = O(1), \nonumber
  \end{align}
  \begin{align}
    \label{eqn:mm3-a5}
    \frac{1}{n} \sum_{i=1}^n \mathrm{Tr}(\mat{V}_i^{-1}\mat{U}_{ij}\mat{V}_i^{-1} & \mat{U}_{ik}\mat{V}_i^{-1}\mat{U}_{il}) = O(1), 
    \qquad \frac{1}{n} \sum_{i=1}^n \mathrm{Tr}(\mat{V}_i^{-1}\mat{U}_{ij}\mat{V}_i^{-1}\mat{U}_{ik})\mathrm{Tr}(\mat{V}_i^{-1}\mat{U}_{il}) = O(1),\\
    & \frac{1}{n} \sum_{i=1}^n \mathrm{Tr}(\mat{V}_i^{-1}\mat{U}_{ij})\mathrm{Tr}(\mat{V}_i^{-1}\mat{U}_{ik})\mathrm{Tr}(\mat{V}_i^{-1}\mat{U}_{il}) = O(1), \nonumber
  \end{align}
  where the determinant $|\mat{V}_i|$ is bounded away from both zero and infinity $\forall i$.

\item \label{hp:mm4} Define $\mat{X}_i^\ast = \left ( \frac{\eta^2_{i\alpha}\mat{V}_i^{-1}}{(1+2\alpha)^{\frac{n_i}{2}+1}} \right )^{\frac{1}{2}}\mat{X}_i$,
for each $i$,  and ${\mat{X}^\ast}=\mbox{Block-Diag}\{\mat{X}_i^\ast : i=1,\ldots, n \}$.
Then the $\mat{X}^\ast$ matrix satisfies
  \begin{equation}
    \label{eqn:mm4}
    \max_{1 \le i \le n} \left [ \frac{({\mat{X}^\ast}^\top \mat{X}^\ast)^{-1}\mat{X}_i^\top \mat{V}_i^{-1}\mat{X}_i}{n} \right ] = O(1).
  \end{equation}
\end{enumerate}


Under these conditions, we can derive the asymptotic distribution of the MDPDE of the parameters in case of linear mixed models
which is presented in the following theorem; the proof is presented in Section SM--2 of the Supplementary Material.

\begin{theorem}\label{lemma:mm-a}
 Consider the setup of the Linear Mixed Model presented in Section \ref{sec:LMM}. 
 Assume that the true data generating density belongs to the model family and 
 that the independent variables satisfy Assumptions \ref{hp:mm1}-\ref{hp:mm4} for a given (fixed) $\alpha\geq 0$. 
 Then, we have the following results as $n\rightarrow\infty$ keeping $n_i$ fixed for each $i$.
  \begin{itemize}
  \item [(i)] There exists a consistent sequence of roots $\hat{\vect{\theta}}_n = (\hat{\vect{\beta}}^\top , \hat{\sigma}^2_j, j = 0,1, \ldots, r)^\top$
  to the minimum DPD estimating equations given in (\ref{eqn:beta-derivative})--(\ref{eqn:sigj-derivative}).
  \item [(ii)] The asymptotic distributions of $\hat{\vect{\beta}}$ and $\hat{\sigma}^2_j$ are independent for all $j=0, 1, \ldots, r$.
  \item [(iii)] The asymptotic distribution of $\mat{\Omega}_n^{-\frac{1}{2}}\mat{\Psi}_n \sqrt{n} (\hat{\vect{\theta}}_n - \vect{\theta})$ is 
  $(k+r+1)$-dimensional normal with mean zero and covariance matrix $\mat{I}_{(k+r+1)}$. 
  In particular, the asymptotic distribution of $({\mat{X}^\ast}^\top \mat{X}^\ast)^{-\frac{1}{2}}({\mat{X}^\prime}^\top \mat{X}^\prime) (\hat{\vect{\beta}} - \vect{\beta})$ 
  is a $k$-dimensional normal with mean zero and covariance matrix $\mat{I}_k$, 
  where  $\mat{X}^\prime$ and $\mat{X}^\ast$ are as defined in Assumptions \ref{hp:mm1} and \ref{hp:mm4}, respectively.
  \end{itemize}
\end{theorem}


\subsection{Influence function}
\label{subsec:lmm-if}

To explore the robustness properties of the coefficient estimates in our treatment of linear mixed models, 
we  derive the influence function of the MDPDEs following the theory explained in \cite{Ghosh2013}.
Denote the density power divergence functional $T_\alpha = (T^{\vect{\beta}}_\alpha, T^{\vect{\Sigma}}_\alpha)$ for 
the parameter vector $\vect{\theta}^\top = (\vect{\beta}^\top, \vect{\Sigma} =(\sigma^2_0,\ldots,\sigma^2_r))$. 
We continue with the notation of the previous subsections. 

The influence function of the estimator $T^{\vect{\beta}}_\alpha$ with contamination at the direction $i_0$ at the point $\vect{t}_{i_0}$ 
is computed to have the form 
\begin{equation}
  \label{eqn:beta-1-if}
  IF_{i_0}(\vect{t}_{i_0},T_\alpha^{\vect{\beta}},G_1,\ldots,G_n) = ({\mat{X}^\prime}^\top \mat{X}^\prime)^{-1} \mat{X}_{i_0}^\top \mat{V}_{i_0}^{-1}(\vect{t}_{i_0} - \mat{X}_{i_0}\vect{\beta})f_{i_0}(\vect{t}_{i_0};\vect{\theta})^\alpha,
\end{equation}
and the corresponding influence function for the estimator $T_\alpha^{\vect{\Sigma}}$ has the form
\begin{equation}
  \label{eqn:sigma-1-if}
  IF_{i_0}(\vect{t}_{i_0},T_\alpha^{\vect{\Sigma}},G_1,\ldots,G_n) =  \left [ \sum_{i=1}^n \frac{\eta_{i\alpha}}{4(1+\alpha)^{\frac{n_i}{2}+2}} T(\mat{V}_i^{-1}\mat{U}_{ij},\mat{V}_i^{-1}\mat{U}_{ik})  \right ]^{-1} \tau_{i_0},
\end{equation}
where 
\begin{equation*}
\label{eqn:tau}
\tau_i = \left ( 
\frac{1}{2} f_i(\vect{t}_i;\vect{\theta})^\alpha [\mathrm{Tr}(\mat{V}_i^{-1}\mat{U}_{ij}) - (\vect{t}_i-\mat{X}_i\vect{\beta})^\top \mat{V}_i^{-1}\mat{U}_{ij}\mat{V}_i^{-1}(\vect{t}_i-\mat{X}_i\vect{\beta})] -\frac{\eta_{i\alpha}\alpha\mathrm{Tr}(\mat{V}_i^{-1}\mat{U}_{ij})}{2(1+\alpha)^{\frac{n_i}{2}+1}}
\right ).
\end{equation*}
The functions $\vect{z} e^{-\vect{z}^\top \vect{z}}$ and $\vect{z}^\top \vect{z} e^{-\vect{z}^\top \vect{z}}$ are bounded for $\vect{z} \in \mathbb{R}^{n_i}$, 
and thus the influence functions in (\ref{eqn:beta-1-if}) and (\ref{eqn:sigma-1-if}) are bounded in $\vect{t}_{i_0}$ for any $i_0$ and any $\alpha >0$. 
For $\alpha = 0$, the influence functions for $T_\alpha^{\vect{\beta}}$ and $T_\alpha^{\vect{\Sigma}}$ are seen to be unbounded; 
indeed this case corresponds to the non-robust maximum likelihood estimator. 
Hence, unlike the MLE, the minimum DPD estimators are B-robust, i.e. their associated influence functions are bounded, for $\alpha >0$. 

Using similar computations, the influence function of the estimators $T_\alpha^{\vect{\beta}}$ and $T_\alpha^{\vect{\Sigma}}$ 
with contamination in all the $n$ cases at the contamination points $\vect{t}_1, \ldots, \vect{t}_n$, respectively, are given by
\begin{equation}
  \label{eqn:beta-all-if}
  IF(\vect{t}_1,\ldots,\vect{t}_n,T_\alpha^{\vect{\beta}},G_1,\ldots,G_n) = ({\mat{X}^\prime}^\top \mat{X}^\prime)^{-1} \sum_{i=1}^n \mat{X}_{i}^\top \mat{V}_{i}^{-1}(\vect{t}_{i} - \mat{X}_{i}\vect{\beta})f_i(\vect{t}_{i_0};\vect{\theta})^\alpha
\end{equation}
and
\begin{equation}
  \label{eqn:sigma-all-if}
IF(\vect{t}_1, \ldots, \vect{t}_n, T_\alpha^{\vect{\Sigma}},G_1,\ldots,G_n) =  \left [ \sum_{i=1}^n \frac{\eta_{i\alpha}}{4(1+\alpha)^{\frac{n_i}{2}+2}} T(\mat{V}_i^{-1}\mat{U}_{ij},\mat{V}_i^{-1}\mat{U}_{ik})\right ]^{-1} \sum_{i=1}^n \tau_{i}.~~~
\end{equation}
In this case also the influence functions are bounded for $\alpha > 0$ and unbounded for $\alpha = 0$.

Now, we compute the sensitivity measures introduced in \citet{Ghosh2013}. For $\alpha > 0$, the gross-error sensitivity and the self-standardized sensitivity of the estimator $T_\alpha^{\vect{\beta}}$ in the case of contamination only in the $i_0$-th direction are given by
\begin{align}
\label{eqn:ges-beta}
\gamma_{i_0}^u(T_\alpha^{\vect{\beta}},G_1,\ldots,G_n) & = \frac{\sup_{\vect{Z}}\{||({\mat{X}^\prime}^\top \mat{X}^\prime)^{-1}\mat{X}_{i_0}\mat{V}_{i_0}^{-\frac{1}{2}}\vect{Z}||e^{-\frac{\vect{Z}^\top \vect{Z}}{2}}\}}{\sqrt{\alpha}(2\pi)^{\frac{n_i\alpha}{2}}|\mat{V}_i|^{\frac{\alpha}{2}}} \\
& = \frac{ \left[\lambda_{max}\left(({\mat{X}^\prime}^\top \mat{X}^\prime)^{-2}\mat{X}_{i_0}^\top \mat{V}_{i_0}^{-1}\mat{X}_{i_0} \right)\right]^{1/2}}{\sqrt{\alpha}(2\pi)^{\frac{n_i\alpha}{2}}|\mat{V}_i|^{\frac{\alpha}{2}} e^{1/2}} \nonumber
\end{align}
and
\begin{align}
\label{eqn:sss-beta}
\gamma_{i_0}^s(T_\alpha^{\vect{\beta}},G_1,\ldots,G_n) & = \frac{\sup_{\vect{Z}} \left \{ \vect{Z}^\top \mat{V}_{i_0}^{-\frac{1}{2}}\mat{X}_{i_0}({\mat{X}^\ast}^\top \mat{X}^\ast)^{-1}\mat{X}_{i_0}^\top \mat{V}_{i_0}^{-\frac{1}{2}} \vect{Z} e^{-\frac{\vect{Z}^\top \vect{Z}}{2}} \right \}^{\frac{1}{2}}}{\sqrt{2\alpha}(2\pi)^{\frac{n_i\alpha}{2}}|\mat{V}_i|^{\frac{\alpha}{2}}} \\
& = \frac{\left[\lambda_{max}\left(({\mat{X}^\ast}^\top \mat{X}^\ast)^{-1}\mat{X}_{i_0}^\top \mat{V}_{i_0}^{-1}\mat{X}_{i_0} \right)\right]^{1/2} }{n \sqrt{\alpha} (2\pi)^{\frac{n_i\alpha}{2}}|\mat{V}_i|^{\frac{\alpha}{2}} e^{1/2}} \nonumber ,
\end{align}
where $\lambda_{max}(\mat{A})$ indicates the largest eigenvalue of the matrix $\mat{A}$, while they are equal to $\infty$ if $\alpha = 0$. 
Details of the computations are provided in Section SM--3 of the Supplementary Material. 
The sensitivity measures for $T_\alpha^{\vect{\Sigma}}$ have no compact form and they are not reported separately.

\subsection{A Particular Example of the LMM}
\label{subsec:particular}

Consider the model defined by Equation (\ref{eqn:lmm}). Here, we study the simplest case in which $n_i = p$,  for all $i= 1, \ldots,n$,
and the associated random effects covariates ($\mat{Z}_{ij}$) are also the same for all $i$. 
In this case, the covariance matrix of $\vect{Y}_i$ is the same for all $i$ and is denoted by $\mat{V}$ having the form
\begin{equation*}
\mat{V}=\mat{V}_i = \sigma_0^2 \left(\mat{I}_p + \sum_{j=1}^r \mat{U}_j \gamma_j\right),
\end{equation*}
where $\gamma_j = \sigma_j^2/\sigma_0^2$ and $\mat{U}_j = \mat{U}_{ij}$ as it is independent of $i$. 
In this situation, we are able to derive an updating expression for the estimation of $\vect{\beta}$, which is very useful for the implementation. 
Consider the objective function rewritten as
\begin{equation*}
  H_n(\vect{\theta}) = \frac{1}{n(2\pi)^{\frac{p\alpha}{2}}|\mat{V}|^{\frac{\alpha}{2}}} \sum_{i=1}^n \bigg [ \frac{1}{(1+\alpha)^{p/2}}-\left (1+\frac{1}{\alpha}\right)e^{-\frac{\alpha}{2}(\vect{Y}_i - \mat{X}_i\vect{\beta})\mat{V}^{-1}(\vect{Y}_i - \mat{X}_i\vect{\beta})}\bigg].
\end{equation*}
Differentiating the above equation with respect to $\vect{\beta}$, Equation (\ref{eqn:beta-derivative}) now corresponds to
\begin{equation*}
  -\sum_{i=1}^n w_i\mat{X}_i^\top \mat{V}^{-1}\vect{Y}_i + \sum_{i=1}^n w_i\mat{X}_i^\top \mat{V}^{-1}\mat{X}_i\vect{\beta} =0,
\end{equation*}
where we denote $w_i = w_i(\vect{\beta}, \sigma_j) = e^{-\frac{\alpha}{2}(\vect{Y}_i - \mat{X}_i\vect{\beta})^\top \mat{V}^{-1}(\vect{Y}_i - \mat{X}_i\vect{\beta})}$. 
Solving for $\vect{\beta}$, we get
\begin{equation*}
  \vect{\beta} = \bigg(\sum_{i=1}^n w_i\mat{X}_i^\top \mat{V}^{-1}\mat{X}_i\bigg)^{-1}\bigg(\sum_{i=1}^n w_i\mat{X}_i^\top \mat{V}^{-1}\vect{Y}_i\bigg),
\end{equation*}
so that, in an iterative fixed point algorithm, the successive iterates have the relation
\begin{equation*}
    \vect{\beta}^{(k+1)} = \bigg(\sum_{i=1}^n w_i(\vect{\beta}^{(k)},\sigma_j^{(k)})\mat{X}_i^\top \mat{V}(\sigma_j^{(k)})^{-1}X_i\bigg)^{-1}\bigg(\sum_{i=1}^n w_i(\vect{\beta}^{(k)},\sigma_j^{(k)})\mat{X}_i^\top \mat{V}(\sigma_j^{(k)})^{-1}\vect{Y}_i\bigg).
\end{equation*}

Note that, the asymptotic distribution of the estimator of $\vect{\beta}$ also has a simpler form. In particular, 
the asymptotic distribution of $\left(\sum\limits_{i=1}^n \mat{X}_i^\top \mat{V}^{-1}\mat{X}_i\right)^{\frac{1}{2}}\sqrt{n}(\hat{\vect{\beta}} - \vect{\beta})$ 
is a $k$-dimensional normal with mean zero and covariance matrix $\upsilon_\alpha^\beta \mat{I}_k$, where 
\begin{equation}
  \label{eqn:beta-asym-var-particular}
  \upsilon_\alpha^\beta = \frac{(1 + \alpha)^{p+2}}{(1 + 2\alpha)^{\frac{p}{2}+1}}.
\end{equation}
Unfortunately, we cannot derive a similar simple form for the estimator of variance parameters.

Furthermore, the following simpler form of the influence function allows us to assess the performance of the sensitivity measure with respect 
to the tuning parameter $\alpha$. The influence function of the functional $T_\alpha^{\vect{\beta}}$ with contamination in the direction $i_0$, 
given in Equation (\ref{eqn:beta-1-if}), can be written as
\begin{equation*}
  IF_{i_0}(\vect{t}_{i_0},T^{\vect{\beta}}_\alpha,G_1,\ldots,G_n) = (1+\alpha)^{\frac{p}{2}+1}\left(\sum_{i=1}^n\mat{X}_i^\top \mat{V}^{-1}\mat{X}_i\right)^{-1}
  \mat{X}_{i_0}^\top \mat{V}^{-1}(\vect{t}_{i_0}-\mat{X}_{i_0}\vect{\beta})w_{i_0}.
\end{equation*}
Using this expression, the gross-error sensitivity for the functional $T_\alpha^{\vect{\beta}}$ is given by
\begin{equation}
  \label{eqn:ges-particular}
  \gamma_{i_0}^u(T_\alpha^{\vect{\beta}})= \frac{(1+\alpha)^{\frac{p}{2}+1}}{\sqrt{\alpha}}
  \left[\lambda_{max}\left(\left[\sum_{i=1}^n \mat{X}_i^\top \mat{V}^{-1} \mat{X}_i\right]^{-2}\mat{X}_{i_0}^\top \mat{V}^{-1}\mat{X}_{i_0} \right)\right]^{1/2} e^{-1/2} .
\end{equation}
Similarly, the self-standardized sensitivity of the functional $T_\alpha^{\vect{\beta}}$ can be written as
\begin{equation}
  \label{eqn:sss-particular}
  \gamma_{i_0}^s (T_\alpha^{\vect{\beta}}) = \frac{(1+\alpha)^{\frac{p+2}{4}}}{n\sqrt{\alpha}} 
  {\left[\lambda_{max}\left(\left[\sum_{i=1}^n \mat{X}_i^\top \mat{V}^{-1} \mat{X}_i\right]^{-1}\mat{X}_{i_0}^\top \mat{V}_{i_0}^{-1}\mat{X}_{i_0} \right)\right]^{1/2}}(2e)^{-1/2}.
\end{equation}
The function $\frac{(1+\alpha)^{\frac{p}{2}+1}}{\sqrt{\alpha}}$ in the gross-error sensitivity (\ref{eqn:ges-particular})
has a minimum for the value $\alpha^\ast = \frac{1}{p+1}$, suggesting that this value of the parameter $\alpha$ gives the most robust estimator. 
Similarly, the function $\frac{(1+\alpha)^{\frac{p+2}{4}}}{\sqrt{\alpha}}$ in the self-standardized sensitivity (\ref{eqn:sss-particular})
has a minimum for the value $\bar{\alpha} = \frac{2}{p}$.
These are in contrast with the previously held knowledge about this parameter which was introduced as a trade-off between efficiency and robustness.

In the following, we present a simple example for which we will compute the theoretical quantities introduced above. 
This example in linear mixed models  has been chosen for its similarity to the case of longitudinal data; 
it is often also named as LMM with random intercept and random slope.

We consider $n=50$ different subjects (groups) and for each of them we have $p=10$ measurements taken with respect to the factor $u_{i2}$, $i=1,\ldots,n$, with two levels, modeled here as a random effect. The $\mat{X}$'s model matrices are simulated from a standard normal. In particular, the model is described by
\begin{equation*}
  \label{eqn:numerical-model}
  \vect{Y}_i = \beta_0 + \beta_1 \mat{X}_i + \vect{u}_{i1} + \vect{u}_{i2}\mat{Z}_{i2}  + \vect{\epsilon}_i,
\end{equation*}
where $u_{i1} \sim N_p(0,\sigma^2_1\mat{I}_p)$, $u_{i2} \sim N(0,\sigma^2_2\mat{I}_2)$ and $\epsilon_i \sim N(0,\sigma_0^2\mat{I}_p)$ and they are independent. 
Hence, for this model, $\vect{\theta} = (\beta_0,\beta_1,\sigma^2_1,\sigma^2_2,\sigma_0^2)$ 
and we take $\vect{\theta} = (1,2,0.25,0.5,0.25)$ as the true values of the parameters. 

\begin{figure}[htbp]
	\begin{center}
		\includegraphics[width=0.35\textwidth]{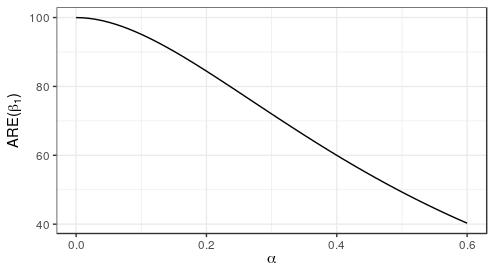}
		\includegraphics[width=0.35\textwidth]{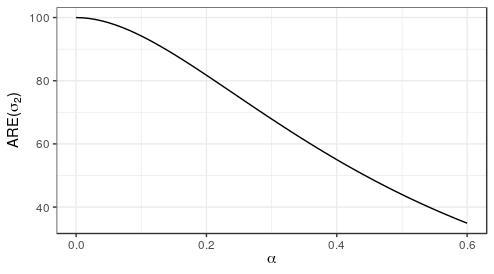}
	\end{center}
	\caption{Asymptotic Relative Efficiency with respect to $\alpha$ for $\beta_1$ and $\sigma^2_2$, respectively.}
	\label{fig:are}
\end{figure}

Using the given values, we compute the variance-covariance matrices $\mat{V}_i$ and the matrices $\mat{\Psi}_n$ and $\mat{\Omega}_n$. 
First, we will look at the \textit{Asymptotic Relative Efficiency} (ARE) of the minimum density power divergence estimators with respect to the fully efficient maximum likelihood estimator. For example, the ARE of $\hat{\vect{\beta}}$ is given by
\begin{equation*}
  ARE(\hat{\vect{\beta}}) = \frac{\upsilon_0^{\vect{\beta}}}{\upsilon_\alpha^{\vect{\beta}}} \times 100
\end{equation*}
where $\upsilon_\alpha^{\vect{\beta}}$ is as defined in Equation (\ref{eqn:beta-asym-var-particular}). The AREs of $\hat{\sigma}^2_i$, $i = 0,1,2$, are similarly defined and they are computed using the general formulation of asymptotic variance.  
Fig. \ref{fig:are} shows the asymptotic relative efficiencies of the estimators of $\beta_1$ and $\sigma^2_1$ for $\alpha \in [0,0.6]$. It is easy to see that there is a loss of efficiency which increases with $\alpha$. 
However, for small positive values of $\alpha$, the estimator retains reasonable efficiency. 
The ARE of the estimators of the other parameters are similar to those displayed here, 
and are given in Section SM--4 of the Supplementary Material.  

\begin{figure}[htbp]
\begin{center}
\subfloat[$\alpha=0$]{
	\includegraphics[width=0.40\textwidth]{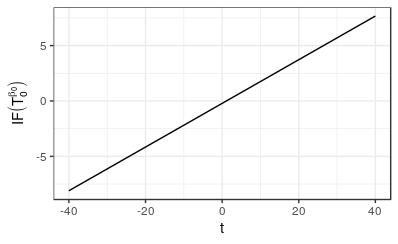}
	\includegraphics[width=0.40\textwidth]{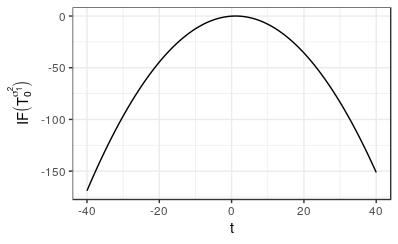}
		\label{fig:a0}} 
	\\
\subfloat[$\alpha=0.05$]{	
\includegraphics[width=0.40\textwidth]{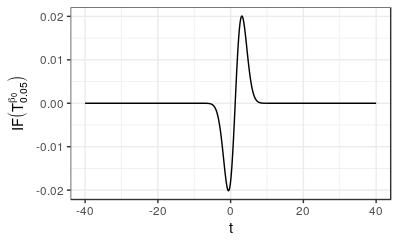}
\includegraphics[width=0.40\textwidth]{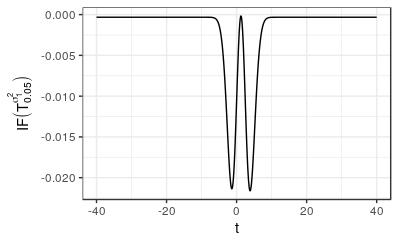}
		\label{fig:a005}} 
\\
\subfloat[$\alpha=0.3$]{	
\includegraphics[width=0.40\textwidth]{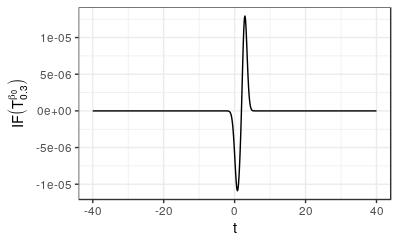}
\includegraphics[width=0.40\textwidth]{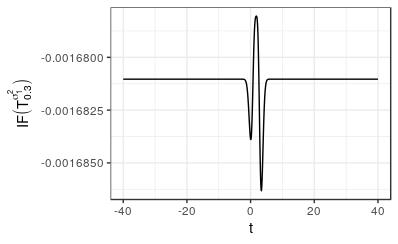}
		\label{fig:a03}} 
\\
\subfloat[$\alpha=0.6$]{	
\includegraphics[width=0.40\textwidth]{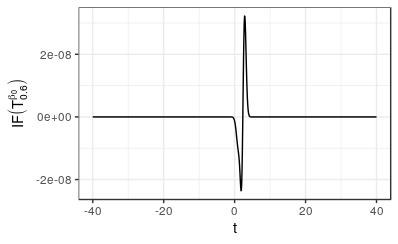}
\includegraphics[width=0.40\textwidth]{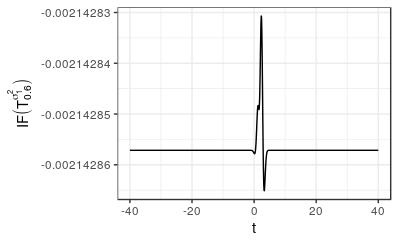}
		\label{fig:a06}} 
\end{center}
\caption{Influence function for the functionals $T_\alpha^{\beta_0}$(left panel) and $T_\alpha^{\sigma_1^2}$ (right panel), 
	for different values of the tuning parameter $\alpha$.}
\label{fig:if-beta0-sigma1}
\end{figure}

On the other hand, to study the robustness properties, Fig. \ref{fig:if-beta0-sigma1} shows the influence functions of $T_\alpha^{\beta_0}$ and $T_\alpha^{\sigma_1^2}$, with respect to $\alpha = 0, 0.05, 0.3, 0.6$. 
Here we have plotted $IF(\vect{t}_1,\ldots,\vect{t}_n,T_\alpha,G_1,\ldots,G_n)$, the influence function of the estimator $T_\alpha$, computed with respect to constant vectors $\underline{\vect{t}} = t  (1,\ldots,1)^\top$ 
for varying $t \in \mathbb{R}$.
Note that, except for the case $\alpha=0$, we can easily see that the influence function is bounded 
as may also be noted from equations (\ref{eqn:beta-1-if}) and (\ref{eqn:sigma-1-if}); 
thus the estimator will be robust with respect to outliers. 
The influence function for the estimators of other parameters behaves similarly; 
these plots are available in Section SM--4 of the Supplementary Material.

Finally, Figure \ref{fig:sensitivity} shows the gross-error sensitivity and the self-standardized sensitivity of the functional $T_\alpha^{\vect{\beta}}$. 
Here, we have considered a particular direction $i_0 \in \{1,\ldots,n\}$. 
Note that, in the present case of balanced data, the choice of $i_0$ does not change the behaviour of the sensitivity measures 
with respect to $\alpha$.

\begin{figure}[h!]
	\begin{center}
		\includegraphics[width=0.49\textwidth]{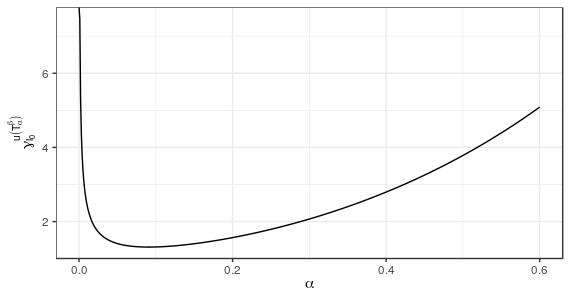}
		\includegraphics[width=0.49\textwidth]{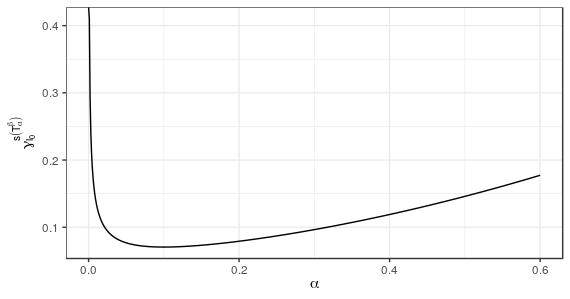}
	\end{center}
	\caption{Gross-error sensitivity (left panel) and self-standardized sensitivity (right panel) of the functional $T_\alpha^{\vect{\beta}}$ with respect to $i_0 = 10$.}
	\label{fig:sensitivity}
\end{figure}

\section{Monte Carlo simulations}
\label{sec:montecarlo}

Here, we describe a simulation study conducted to assess the performance of the proposed estimator in case of LMMs. 
It will be compared to the primary existing competitors both under pure data as well as under contaminated data.

\subsection{Model setting}
\label{subsec:setting}

This model setting has been introduced in \citet{Agostinelli2016} and reported here in order to facilitate the comparison of the considered estimators.

Consider an LMM for a 2-way cross classification with interaction, where the model is given by
\begin{equation*}
  \label{eqn:2waylmm}
  y_{fgh} = \vect{x}^\top_{fgh}\vect{\beta}_0 + a_f + b_g + c_{fg} + e_{fgh},
\end{equation*}
where $f=1,\ldots,F, g= 1, \ldots ,G$, and $h=1,\ldots,H$. 
Here, we set $F=2, G=2$ and $H=3$ getting $p=F\times G \times H= 12$.  
Also $\vect{x}_{fgh}$ is a $k \times 1$ vector where the last $k - 1$ components are from a standard multivariate normal 
and the first component is identically equal to $1$, 
and $\vect{\beta}_{0}=(0,2,2,2,2,2)^{\top}$ is a $k \times 1$ vector of the fixed parameters with $k=6$. 
The random variables $a_{f}$, $b_{g}$ and $c_{fg}$ are the random effects 
which are normally distributed with variances $\sigma_{a}^{2}$, $\sigma_{b}^{2}$, and $\sigma_{c}^{2}$. 
Arranging the $y_{fgh}$ in lexicon order (ordered by $h$ within $g$ within $f$) we obtain the vector $\vect{y}$ of dimension $p$ 
and in the similar way the $p\times k$ matrix $\mat{x}$ obtained arranging $\vect{x}_{fgh}$. 
Similarly, we set $\vect{a}=(a_{1},\ldots,a_{F})^{\top}$, $\vect{b}=(b_{1},\ldots,b_{G})^{\top}$ and $\vect{c}=(c_{11},\ldots,c_{FG})^{\top}$, 
that is, $\vect{a}\sim N_{F}(\vect{0},\sigma_{a}^{2}\mat{I}_{F})$ and similarly for $\vect{b}$ and $\vect{c}$, 
while $\vect{e}=(e_{111},\ldots,e_{FGH})^{\top}\sim N_{p}(\vect{0},\sigma_{e}^{2}\mat{I}_{p})$. 
Hence $\vect{y}$ is a $p$ multivariate normal with mean $\vect{\mu}=\mat{x}\vect{\beta}$ and variance matrix 
$\mat{\Sigma}_{0}=\mat{\Sigma}(\eta_{0},\vect{\gamma}_{0}) =\eta_{0}(\mat{V}_{0}+\sum_{j=0}^{J}\gamma_{j}\mat{V}_{j})$, 
where $\mat{V}_{0}=\mat{I}_{p}$, $\mat{V}_{1}=\mat{I}_{F}\otimes\mat{J}_{G}\otimes\mat{J}_{H}$, 
$\mat{V}_{2}=\mat{J}_{F}\otimes\mat{I}_{G}\otimes \mat{J}_{H}$, 
and $\mat{V}_{3}=\mat{I}_{F}\otimes \mat{I}_{G}\otimes\mat{J}_{H}$; 
$\otimes$ is the Kronecker product and $\mat{J}_k$ is a $k \times k$ matrix of ones. 
We took $\sigma_{a}^{2}=\sigma_{b}^{2}=1/16$ and $\sigma_{c}^{2}=1/8$. 
Then $\vect{\gamma_0}=(\gamma_{01},\gamma_{02},\gamma_{03})^{\top}=(\sigma_{a}^{2}/\sigma_{e}^{2},\sigma_{b}^{2}/\sigma_{e}^{2}, \sigma_{c}^{2}/\sigma_{e}^{2})^{\top}=(1/4,1/4,1/2)^{\top}$ and $\eta_{0}=\sigma_{e}^{2}=1/4$. 
We consider a sample of size $n=100$ and four levels of contamination $\varepsilon=0,5,10$ and $15\%$. 
Hence, $n\times\varepsilon$ observations are contaminated by replacing $n\times\varepsilon$ elements of the vector $\vect{y}$ 
by observations from $\vect{y}_{0}\sim N_{p}(\mat{x}_{0}\vect{\beta}_{0}+\vect{\omega}_{0},\mat{\Sigma})$ 
and the corresponding components of $\mat{x}$ are replaced by the components of $\mat{x}_{0}$. 
The first column of $\mat{x}_{0}$ is identically equal to $1$ 
while the last $k - 1$ columns are from $N_{p\times (k-1)}(\vect{\phi}_{0},0.005^{2}\mat{I}_{p\times (k-1)})$ 
and all the components of $\vect{\phi}_{0}$ equal to $1$ in the case of low leverage outliers (lev1) or to $20$ for large leverage outliers (lev20). 
$\vect{\omega}_{0}$ is a $p$-vector of constants all equal to $\omega_{0}$ taken in a grid of values 
which would generate unlikely responses for the model and allow us to explore the behavior of our estimator under such adverse conditions.

For each combination of these factors we compute the CVFS-estimator described in \citet{Copt2006} 
with Rocke $\rho$ function and with asymptotic rejection probability set to $0.01$ 
as implemented in the \texttt{R} \citep{cran} package \texttt{robustvarComp} \citep{Agostinelli2019}, 
the SMDM estimator introduced by \citet{Koller2013} as implemented in the \texttt{R} package \texttt{robustlmm} \citep{Koller2016}, 
and our proposed MDPDE with different choices of $\alpha$ in  $\{0, 0.01, 0.1, 0.2, \ldots,1\}$; 
note that $\alpha^\ast = 1/(p+1) = 1/13$ and $\bar{\alpha} = 2/p = 1/6$. For each case we run $500$ Monte Carlo replications.

\subsection{Performance Measures}

Let $(\vect{y},\mat{x})$ be an observation independent of the sample 
$(\vect{y}_{1}, \mat{x}_{1}), \ldots,(\vect{y}_{n}, \mat{x}_{n})$ used to compute $\widehat{\vect{\beta}}$ 
and let $\widehat{\vect{y}}=\mat{x} \widehat{\vect{\beta}}$ be the predicted value of $\vect{y}$ using $\mat{x}$.
Then, the squared Mahalanobis distance between $\widehat{\vect{y}}$ and $\vect{y}$ using the matrix $\mat{\Sigma}_{0}$ is
\begin{align*}
m(\widehat{\vect{y}}, \vect{y}, \mat{\Sigma}_{0}) & = (\widehat{\vect{y}}-\vect{y})^{\top} \mat{\Sigma}_{0}^{-1}(\widehat{\vect{y}}-\vect{y}) 
= (\widehat{\vect{\beta}}-\vect{\beta}_{0})^{\top} \mat{x}^{\top} \mat{\Sigma}_{0}^{-1} \mat{x} (\widehat{\vect{\beta}}-\vect{\beta}_{0})
+ (\vect{y}-\mat{x} \vect{\beta}_{0})^{\top} \mat{\Sigma}_{0}^{-1} (\vect{y}-\mat{x} \vect{\beta}_{0}).
\end{align*}
Since $\vect{y}-\mat{x}\vect{\beta}_{0}$ is independent of $\mat{x} $ and has covariance matrix $\mat{\Sigma}_{0}$, putting $\mat{A} = \mathbb{E}(\mat{x}^{\top} \mat{\Sigma}_{0}^{-1} \mat{x})$ we have
\begin{align*}
\mathbb{E} \left[ m(\widehat{\vect{y}}, \vect{y}, \mat{\Sigma}_{0})\right] & = \mathbb{E}\left[  (\widehat{\vect{\beta}} - \vect{\beta}_{0})^{\top} \mat{A} (\widehat{\vect{\beta}} - \vect{\beta}_{0}) \right] 
+ \operatorname{trace} \left[\mat{\Sigma}_{0}^{-1} (\vect{y}-\mat{x} \vect{\beta}_{0})
(\vect{y}- \mat{x} \vect{\beta}_{0})^{\top}\right]\\
& = \mathbb{E}\left[ (\widehat{\vect{\beta}}-\vect{\beta}_{0})^{\top} \mat{A}(\widehat{\vect{\beta}}-\vect{\beta}_{0})\right] + p.
\end{align*}
Then, to evaluate an estimator $\widehat{\vect{\beta}}$ of $\vect{\beta}$ by its prediction performance we can use
\begin{equation*}
\mathbb{E}\left[ m(\widehat{\vect{\beta}},\vect{\beta}_{0},\mat{A}) \right] = \mathbb{E} \left[  (\widehat{\vect{\beta}} - \vect{\beta}_{0})^{\top} \mat{A}(\widehat{\vect{\beta}}-\vect{\beta}_{0})\right] .
\end{equation*}
Let $N$ be the number of replications in the simulation study, and let $\widehat{\vect{\beta}}_{j}$, $1\leq j\leq N$ be the value of $\widehat{\vect{\beta}}$ at the $j$-th replication, 
then we can estimate $\mathbb{E}\left[m(\widehat{\vect{\beta}},\vect{\beta}_{0},\mat{A}) \right]$ 
by the mean square Mahalanobis distance as
\begin{equation*}
\text{MSMD} = \frac{1}{N} \sum_{j=1}^{N} m(\widehat{\vect{\beta}}_{j}, \vect{\beta}_{0},\mat{A}).
\end{equation*}
It is easy to prove that, as in this case, $\mat{x}$ is a $p\times k$ matrix where the cells are independent $N(0,1)$ random variables, then $\mat{A} = \operatorname{trace}(\mat{\Sigma}_{0}^{-1}) \mat{I}_{k}$.

Given two $p$-dimensional covariance matrices $\mat{\Sigma}_{1}$ and $\mat{\Sigma}_{0}$, 
one way to measure how close $\mat{\Sigma}_{1}$ and $\mat{\Sigma}_{0}$ are is through the use of the Kullback-Leibler divergence 
between two multivariate normal distributions with the same mean and covariance matrices equal to 
$\mat{\Sigma}_{1}$ and $\mat{\Sigma}_{0}$, given by
\begin{equation*}
\text{KLD}(\mat{\Sigma}_{1},\mat{\Sigma}_{0})=\text{trace} \left(  \mat{\Sigma}_{1}\mat{\Sigma} _{0}^{-1}\right) -\log\left(  \det(\mat{\Sigma}_{1}\mat{\Sigma}_{0}^{-1})\right)- p .
\end{equation*}
Since $(\eta_{0},\vect{\gamma}_{0})$ determines $\mat{\Sigma}_{0}=\mat{\Sigma}(\eta_{0},\vect{\gamma}_{0})$, 
the covariance matrix of $\vect{y}$ given $\mat{x}$ for the particular LMM considered in our simulation (as described in Section \ref{subsec:setting}), one way to measure the performance of an estimator 
$(\widehat{\eta},\widehat{\vect{\gamma}})$ of $(\eta_{0},\vect{\gamma}_{0})$ is by
$\mathbb{E}\left[ \text{KLD}(\mat{\Sigma}(\widehat{\eta},\widehat{\vect{\gamma}}),\mat{\Sigma}_{0})\right]$.
Let $(\widehat{\eta}_{j},\widehat{\vect{\gamma}}_{j}),1\leq j\leq N$, 
be the value of $(\widehat{\eta},\widehat{\vect{\gamma}})$ at the $j$-th replication, 
then we can estimate $\mathbb{E}\left[\text{KLD}(\mat{\Sigma}(\widehat{\eta},\widehat{\vect{\gamma}}),\mat{\Sigma}_{0})\right]$ 
by the mean Kullback-Leibler divergence
\begin{equation*}
\text{MKLD}=\frac{1}{N}\sum_{j=1}^{N}\text{KLD}(\mat{\Sigma}(\widehat{\eta}_{j},\widehat{\vect{\gamma}}_{j}),\mat{\Sigma}_{0}).
\end{equation*}

\subsection{Results}

We begin with the performance of the estimators in the absence of contamination. Table \ref{tab:rel-eff} shows 
the relative efficiency of the CVFS-estimator, the SMDM-estimator, and the MDPDE for different values of $\alpha$  
with respect to maximum likelihood. 
The efficiency of the estimators of $\vect{\beta}$ has been measured by the MSMD ratio while 
the MKLD ratio was used for the efficiency of an estimator of $(\eta,\vect{\gamma})$.   
\begin{table}[h]
\centering
\begin{tabular}{rr|rr}
   \hline
Method & ($\alpha$) & MSMD EFF. & MKLD EFF. \\
   \hline
SMDM   & -- &   0.956  &  0.147 \\
CVFS   & -- &   0.706  &  0.453 \\
MDPDE & 0.01 &   0.999  &  0.996 \\
  & $\alpha^\ast$ &   0.960  &  0.945 \\
  & 0.1 &   0.937  &  0.915 \\
  & $\bar{\alpha}$ &   0.853  &  0.814 \\
  & 0.2 &   0.805  &  0.760 \\
  & 0.3 &   0.658  &  0.603 \\
  & 0.4 &   0.519  &  0.470 \\
  & 0.5 &   0.400  &  0.361 \\
  & 0.6 &   0.302  &  0.273 \\
\hline
\end{tabular}
\caption{Relative efficiency for the SMDM-estimator, CVFS-estimator and MDPDE for different values of $\alpha$ with respect to the maximum likelihood computed by the MSMD for the fixed terms $\vect{\beta}$ and by the MKLD for the random terms.}
\label{tab:rel-eff}
\end{table}

The MDPDEs exhibit a high relative efficiency, even greater than the competitor estimators, for small values of $\alpha$, 
while the efficiency decreases with increasing $\alpha$. 
Note that the MDPDEs are far more successful in retaining the efficiency of the estimators of the random component. 
For very small values of $\alpha$ the MDPDEs dominate either competitor 
(at least up to $\alpha = \alpha^*$ for SMDM, and at least up to $\alpha = 0.2$ for CVFS) in terms of both (MSMD and MKLD) efficiency measures.  
As the value of $\alpha$ increases, the MSMD efficiency of the MDPDE eventually lags behind its competitors, 
but in terms of MKLD efficiency it beats both competitors at least up to $\alpha = 0.4$. 
On the whole it is clear that under pure data, a properly chosen member of the MDPDE class can perform competitively, 
if not better, compared to the SMDM and CVFS estimators. 

Now, we consider the contamination setting. The Figures presented in Section SM--5 of the Supplementary Material show the MSMD and the MKLD of the MDPDE for different values of $\alpha$ compared to the CVFS- and SMDM-estimators, as a function of $\omega_0$.

For a simpler comparison, Table \ref{tab:max-10} reports the maximum values of MSMD and MKLD over the values of $\omega_0$ considered in the range of our Monte Carlo setting. 

\begin{table}[htbp]
	\centering
	\begin{tabular}{rr |rr|rr}
		\hline
		& &	\multicolumn{2}{c|}{MSMD} & 	\multicolumn{2}{c }{MKLD}\\
		Method & ($\alpha$) & lev1 & lev20 & lev1 & lev20\\
		\hline
		CVFS	& - &	0.010	&	0.122	&	0.197	&	1.057	\\
		SMDM	& - &	0.021	&	0.450	&	0.617	&	7.993	\\
                MDPDE   & 0	&	9.007	&	9.005	&	3.508e22	&	1.605e25	\\
		  &  0.01	&	1.114	&	0.120	&	106.185	&	0.732	\\
		 & $\alpha^\ast$	&	0.017	&	0.121	&	0.650	&	0.716	\\
		  & 0.1	&	0.012	&	0.121	&	0.387	&	0.710	\\
		 & $\bar{\alpha}$	&	0.008	&	0.122	&	0.139	&	0.695	\\
		 & 0.2	&	0.008	&	0.122	&	0.105	&	0.688	\\
		 & 0.3	&	0.007	&	0.123	&	0.106	&	0.673	\\
		 & 0.4	&	0.007	&	0.125	&	0.116	&	0.665	\\
		 & 0.5	&	0.008	&	0.127	&	0.137	&	0.662	\\
		 & 0.6	&	0.010	&	0.130	&	0.170	&	0.666	\\
		\hline
	\end{tabular}
	\caption{Maximum values of MSMD and MKLD for the CVFS-, SMDM-estimators and for the MDPDE at different values of $\alpha$ under $10\%$ of outlier contamination.}
	\label{tab:max-10}
\end{table}

Small values of $\alpha$, as expected, provide much higher maximum values with respect to the other estimators in Table \ref{tab:max-10}.      
However for slightly larger values of $\alpha$, the MDPDEs are extremely competitive with the existing estimators. 
It may be easily observed that the MDPDE at $\bar{\alpha}$ clearly beats both competitors (CVFS and SMDM) 
over both performance measures at both leverage values (except at MSMD, lev20, where its performance measure is equal to that of CVFS). In this example, the MDPDE at $\alpha = 0.2$ fares even better. 
The values $\alpha^*$ and $\bar{\alpha}$ represent theoretical optimal choices, 
even though they may not present the lowest maximum values of MSMD or MKLD measures.  

\begin{figure}[h]
	\centering
	\subfloat[MSMD performance of the estimators of $\vect{\beta}$]{
		\includegraphics[width=0.5\textwidth]{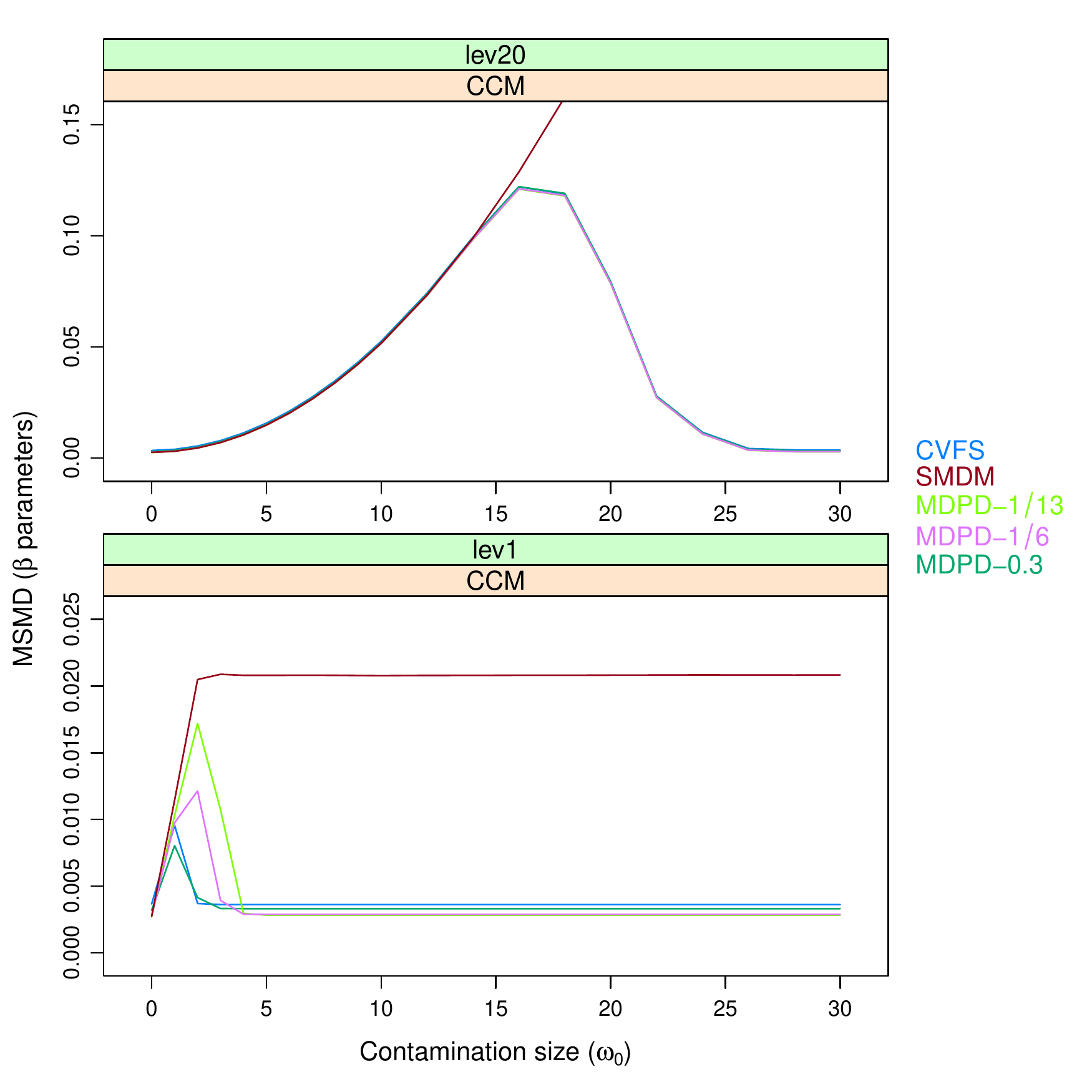}
		\label{fig:msmd-10-selected}} ~
	\subfloat[MKLD performance of the estimators of $(\eta,\vect{\gamma})$]{
		\includegraphics[width=0.5\textwidth]{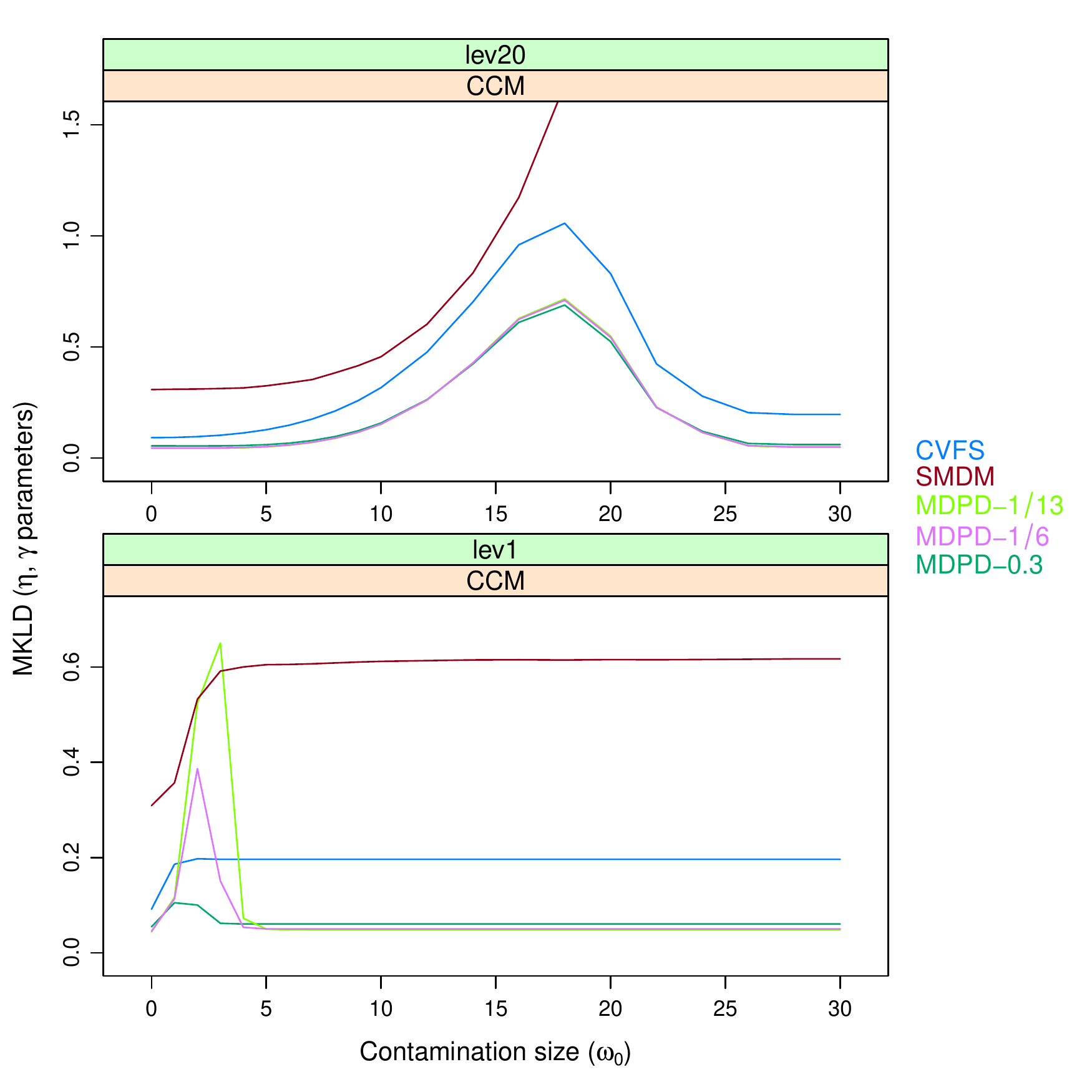}
		\label{fig:mkld-10-selected}} ~
	\caption{Performance of the MDPD-estimators of $\vect{\beta}$ and $(\eta,\vect{\gamma})$ 
		for $\alpha = \alpha^\ast, \bar{\alpha},0.3$, compared to the CVFS- and SMDM-estimators, under $10\%$  outlier contamination.}
	\label{fig:sim2}
\end{figure}

 Figures \ref{fig:msmd-10-selected} and \ref{fig:mkld-10-selected} display the MSMD and MKLD as function of $\omega_0$, comparing the CVFS- and SMDM-estimators with the MDPDEs for three chosen values of $\alpha$, under 10\% of outlier contamination. In particular, we choose $\alpha^\ast$ and $\bar{\alpha}$ since they are the values suggested by theory, and $\alpha=0.3$ since it shows the lowest (or very close to the lowest) maximum values of MSMD and MKLD.
We can see that most of the MDPDEs outperform the CVFS- and SMDM-estimators, especially in case of leverage 20 (lev20), 
where the SMDM-estimator shows an unbounded behaviour. On the other hand, in the case of leverage 1 (lev1), 
even if the CVFS-estimator presents lower maximum value of MSMD and MKLD for very small values of $\omega_0$, 
the MDPDEs show a better performance when $\omega_0$ increases.
In fact the MDPDE at $\alpha = 0.3$ is competitive or better than CVFS at all values of $\omega_0$.  

\section{Real-data example: Extrafoveal Vision Acuity}
\label{sec:example}

Let us now present an application of the proposed estimation method to a real data example. 
We compare the estimates obtained by the minimum DPD method with those obtained using the classical (non-robust) restricted MLE, 
computed using the \texttt{lmer} function in R, as well as the robust competitors, the SMDM-estimator and the CVFS-estimator. 
A very important consideration in real situations is the selection of an ``optimum'' value of $\alpha$ that applies to the given data set. 
We will use different values of $\alpha$ to highlight the behavior of the estimator seen in the simulations. 
In general, we are going to consider the values $\alpha^\ast$ and $\bar{\alpha}$, derived from theoretical computations, 
as suggested optimal values.

We consider the study conducted by \citet{Fromer2015} about the relationship between individual differences in foveal visual acuity and extrafoveal vision (acuity and crowding) and reading time measures, such as reading rate and preview benefit.

There were 40 participants in the study, with normal visual acuity measured with the adaptive computerized Freiburg Acuity Test (FrACT) \citep{Bach1996}. The study was organized in two test sessions. During the first session, the extrafoveal vision assessment (EVA) was provided, involving a test of crowded and uncrowded extrafoveal vision. In addition, visual acuity of fovea was measured using the FrACT. The second session was taken after a week, consisting of an eye-tracking experiment with list reading followed by the EVA procedure. The EVA was performed considering four test conditions: identification of single letters and flanked letters in the left and right visual field. Here, we consider only data coming from measurements with the EVA procedure and do not deal with data related to the reading task.

This kind of data can be modeled using a Linear Mixed Model. In particular, we studied a repeated measures \textit{Analysis of Variance} (ANOVA) of the threshold eccentricities (TE) with random effects given by extrafoveal vision (EV)(single versus crowded letter), hemifield (H)(left, right), and test repetition ($T_1$, $T_2$, $T_3$). Thus, combining the factors given above, we have $p=12$ measurements for each subject (participant). 
The model for each subject ($i$-th) has the form
\begin{align*}
  TE_i = &\beta_0 + \beta_1 EV_i + \beta_2 H_i +\beta_3 T2\_1_i + \beta_4 T3\_2_i + \beta_5 (EV*H)_i + \beta_6 (EV*T2\_1)_i + \\
  & \beta_7 (EV*T3\_3)_i + \beta_8 (H*T2\_1)_i + \beta_9 (H*T3\_2)_i + \beta_{10} (EV*H*T2\_1)_i + \\
  & \beta_{11} (EV*H*T3\_2)_i + u_1 + EV u_2 + H u_3 + T2\_1 u_4 + T3\_2 u_5 + (EV*H) u_6 + \\
  & (EV*T2\_1) u_7 + (EV*T3\_3) u_8 + (H*T2\_1) u_9 + (H*T3\_2) u_{10} + \epsilon_i
\end{align*}
where $i = 1, \ldots, n$, $n = 40$, while $T2\_1$ and $T3\_2$ substitute the factor time $(T_1,T_2,T_3)$ indicating the transitions between the first and second sessions, and between third and second sessions, respectively. Hence, we have 12 fixed effect parameters $(\beta_0, \ldots, \beta_{11})$, and 10 random effects of which we will estimate the variance components $\sigma^2_j$, $j = 1, \ldots, 10$.

Complete results of the present study are reported in Section SM--6 of the Supplementary Material. 
In particular, Table 3 of the Supplementary Material shows the estimates of model parameters obtained using the \texttt{lmer} estimator, the SMDM- and CVFS-estimators, the MDPDE for different values of $\alpha$.
For increasing $\alpha$, the MDPDE's capacity to accurately estimate the variance components drop, especially for $\alpha \ge 0.4$, 
while the estimates of the fixed terms do not significantly change. 
It may be seen that the SMDS-estimator has a poor performance. 
On the other hand, the MDPDEs for $\alpha=0.05$ and $\alpha=1/13$ 
show similar estimates to those obtained using \texttt{lmer} and the CVFS-estimator.

Finally, we tested the \texttt{lmer} estimator, the CVFS-estimator and the MDPDE with $\alpha = 1/13$ in the case where some TE values are substituted by outlying values. In particular, we implemented an iterative procedure where, in each step, an outlying observation is added. Let $\mat{X}$ be the $(40 \times 12)$ matrix of the TE values. After selecting a random cell $(i,j)$, with $i \in \{1,\ldots,40\}$ and $j \in \{1,\ldots,12\}$, $X_{ij}$ is replaced by a value sampled from $N(kv_j, 0.1^2)$, where $k = 10$, $\vect{v}$ is the eigenvector corresponding to the smallest eigenvalue of the maximum likelihood estimate of the covariance matrix and ${v}_j$ indicates the $j$-th component of the vector $\vect{v}$. Before adding the next outlying value, the estimates of \texttt{lmer}, the CVFS-estimator and MDPDE with $\alpha =1/13$ are computed. We repeated the procedure until 9 values had been substituted. 
In the Supplementary Material, Tables 4, 5 and 6 show the estimates and the corresponding $p$-values obtained using the usual \texttt{lmer}, the CVFS-estimator and the proposed MDPDE, respectively, as the number of substituted cells ($m$) increases.

\begin{table}[htbp]
  \centering
  \resizebox{\textwidth}{!}{
    \begin{tabular}{rr|rrrrrrrrrrrr}
      && Intercept & EV & H & T2\_1 & T3\_2 & EV*H & EV*T2\_1 & EV*T3\_2 & H*T2\_1 & H*T3\_2 & EV*H*T2\_1 & EV*H*T3\_2 \\
      \hline
  \texttt{lmer} & $m=0$ &0.000 & 0.000 & 0.685 & 0.200 & 0.126 & 0.000 & 0.042 & 0.066 & 0.103 & 0.083 & 0.139 & 0.007  \\
  &$m = 9$ &0.000 & 0.000 & 0.383 & 0.685 & 0.844 & 0.011 & 0.881 & 0.816 & 0.734 & 0.789 & 0.858 & 0.486 \\
  \hline
  CVFS & $m=0$ &0.000 & 0.000 & 0.987 & 0.777 & 0.565 & 0.000 & 0.639 & 0.312 & 0.188 & 0.318 & 0.336 & 0.038 \\
  &$m=9$ & 0.000 & 0.000 & 0.978 & 0.932 & 0.763 & 0.040 & 0.827 & 0.831 & 0.686 & 0.694 & 0.724 & 0.669 \\
  \hline
  MDPDE 1/13&$m=0$ & 0.000 & 0.000 & 0.977 & 0.716 & 0.534 & 0.000 & 0.493 & 0.263 & 0.139 & 0.189 & 0.252 & 0.027 \\
   & $m=9$ & 0.000 & 0.000 & 0.674 & 0.903 & 0.438 & 0.000 & 0.724 & 0.347 & 0.075 & 0.217 & 0.193 & 0.009 \\
  
  \hline
\end{tabular}}
\caption{$p$-values obtained from the estimates of the \texttt{lmer} estimator, the CVFS-estimator and the MDPDE with $\alpha = 1/13$, for uncontaminated data ($m=0$) and for the data set with $m=9$ substituted cells. }
\label{tab:summarized-results}
\end{table}

Here, we summarize the obtained results in Table \ref{tab:summarized-results}, which reports the $p$-values of the tests, checking whether the parameters are significantly different from zero giving an idea of the importance of the corresponding variables, for uncontaminated data and when $m=9$ cells are substituted.
  In Tables 4, 5 and 6 of the Supplementary Material, the estimates obtained using \texttt{lmer} are more affected than those given by the CVFS-estimator and the MDPDE. On the other hand, the MDPDE seems quite stable with respect to the corresponding $p$-values, while those computed using \texttt{lmer} and the CVFS-estimator show large variations.

\section{Conclusions}
\label{sec:conclusions}

In this paper, we have developed an estimator based on the density power divergences to deal with the robustness issues 
in the linear mixed model setup. 
We demonstrated that this MDPDE satisfies the desirable properties of an estimator, such as consistency and asymptotic normality. 
In order to assess the robustness properties, the influence function and sensitivity measures of the estimator were computed. 
We found that the estimator is B-robust for $\alpha > 0$. 
From a practical point of view, the choice of the value of the tuning parameter $\alpha$ is fundamental in applications. 
The behaviour of the sensitivity measures suggested two optimal values, denoted by $\alpha^\ast$ and $\bar{\alpha}$, 
depending on the dimension $p$, where the term ``optimal'' is in the sense of providing minimum sensitivity, and thus producing maximum robustness. 
The existence of such values is in contrast to the previous knowledge about the parameter $\alpha$. 
Indeed, it was shown that when $\alpha$ continues to increase, beyond a certain value we lose both robustness and efficiency.

The simulation study confirmed how the performance of the minimum density power divergence estimator changes with respect to $\alpha$. 
Furthermore, the MDPDE outperforms the competitor estimators; 
indeed our approach leads to more resistant estimators in the presence of case-wise contamination. 
Finally, the application of our estimator to a real-life data set indicated that the MDPDE has similar results 
to the classical maximum likelihood estimator, with the advantage of being resistant to the presence of few random cell-wise outliers.

We feel that many important extensions of this work are necessary and can be potentially useful. 
So far, the MDPDE has been implemented only for balanced data (although the theory that we have developed is perfectly general). 
In future, we propose to extend the implementation to the more general case of groups with possibly different dimensions.
Also, the linear mixed models are based on normality assumptions, 
it would be useful to extend the application of the MDPDE to the larger class of generalized linear mixed models. 
The problem of testing of hypothesis also deserves a deeper look in the linear mixed models scenario.

\section*{Supplementary Material}

  The Supplementary Material contains the assumptions needed to prove the asymptotic normality of the estimator in Section SM--1, while the proof of Theorem \ref{lemma:mm-a} is reported in Section SM--2. Section SM--3 shows the simplification of the sensitivity measures in case of balanced data. In Section SM--4 we report the missing plots about theoretical quantities. Finally, further results obtained from the Monte Carlo experiments are presented in Section SM--5, whereas Section SM--6 contains complete results from the study of the real-data example.

\bibliography{dpd-lmm}

\end{document}


\title{Supplementary Material for ``Robust estimation under Linear Mixed Models: Minimum Density Power Divergence approach''}
\date{}
\author[1]{Giovanni Saraceno}
\author[2]{Abhik Ghosh}
\author[2]{Ayanendranath Basu}
\author[1]{Claudio Agostinelli}
\affil[1]{Department of Mathematics, University of Trento, Trento, Italy}
\affil[2]{Interdisciplinary Statistical Research Unit, Indian Statistical Institute, Kolkata, India}
\maketitle

\section*{Supplementary Material}

The Supplementary Material contains the assumptions needed to prove the asymptotic normality of the estimator in Section \ref{sm:sec:asymptotic}, while the proof of Theorem $1$ of the main paper is reported in Section \ref{sm:sec:proof-lemma}. Section \ref{sm:sec:sensitivity} shows the simplification of the sensitivity measures in case of balanced data. In Section \ref{sm:sec:numerical-study} we report the missing plots about theoretical quantities. Finally, further results obtained from the Monte Carlo experiments are presented in Section \ref{sm:sec:montecarlo}, whereas Section \ref{sm:sec:example} contains complete results from the study of the real-data example.

\section{Asymptotic properties}
\label{sm:sec:asymptotic}

Continuing with the notation and set-up of Section 2 of the main paper, let  $\mathcal{F}_{i,\vect{\theta}}$ be the parametric model. 
Assume that there exists a best fitting parameter of $\vect{\theta}$ which is independent of the index $i$ of the different densities 
and let us denote it by $\vect{\theta}^g$. We also assume that all the true densities $g_i$ belong to the model family, 
i.e. $g_i = f_i(\cdot;\vect{\theta})$ for some common $\vect{\theta}$, and in that case the best fitting parameter is nothing but the true parameter $\vect{\theta}$. 
We define, for each $i = 1, \ldots, n$, the $p \times p$ matrix $\mat{J}^{(i)}$ as
\begin{align*}
  \mat{J}^{(i)} =& (1+\alpha)\left[\int u_i(\vect{y};\vect{\theta}^g)u^\top_i(\vect{y};\vect{\theta}^g)f_i^{1+\alpha}(\vect{y};\vect{\theta}^g) dy \nonumber \right. \\
  & \left. - \int \{ \nabla u_i(\vect{y};\vect{\theta}^g) + \alpha u_i(\vect{y};\vect{\theta}^g)u^\top_i(\vect{y};\vect{\theta}^g)\}\{g_i(\vect{y})-f_i(\vect{y};\vect{\theta}^g)\} f_i(\vect{y};\vect{\theta}^g)^\alpha dy \right].
  \label{eqn:j-matrix}
\end{align*}
We further define the quantities
\begin{equation*}
  \mat{\Psi}_n = \frac{1}{n} \sum_{i=1}^n \mat{J}^{(i)},
  \label{eqn:psi-matrix}
\end{equation*}
and
\begin{equation*}
  \mat{\Omega}_n = (1+\alpha)^2\frac{1}{n}\sum_{i=1}^n \left[ \int u_i(\vect{y};\vect{\theta}^g)u^\top_i(\vect{y};\vect{\theta}^g)f_i(\vect{y};\vect{\theta}^g)^{2\alpha}g_i(\vect{y})dy - \vect{\xi}_i \vect{\xi}^\top_i\right],
  \label{eqn:omega-matrix}
\end{equation*}
where
\begin{equation*}
  \vect{\xi}_i = \int u_i(\vect{y};\vect{\theta}^g)f_i(\vect{y};\vect{\theta}^g)^\alpha g_i(\vect{y})dy.
  \label{eqn:xi-vect}
\end{equation*}

We will make the following assumptions to establish the asymptotic properties of the MDPDE
under this general set-up of independent but non-homogeneous observations \citep{Ghosh2013}.

\begin{enumerate}[label=\text{(A\arabic*)}]
\item \label{hp:a1} The support $\chi = \{\vect{y} | f_i(\vect{y};\vect{\theta}) > 0\}$ is independent of $i$ and $\vect{\theta}$ for all $i$; the true distributions $G_i$ are also supported for all $i$. 
\item \label{hp:a2} There is an open subset $\omega$ of the parameter space $\Theta$, containing the best fitting parameter $\vect{\theta}^g$ such that for almost all $\vect{y} \in \chi$, and all $\vect{\theta} \in \Theta$, all $i=1, \ldots, n$, the density $f_i(\vect{y};\vect{\theta})$ is thrice differentiable with respect to $\vect{\theta}$ and the third partial derivatives are continuous with respect to $\vect{\theta}$. 
\item \label{hp:a3} For $i=1,\ldots,n$, the integrals $\int f_i(\vect{y};\vect{\theta})^{1+\alpha}dy$ and $\int f_i(\vect{y};\vect{\theta})^\alpha g_i(\vect{y})dy$ can be differentiated thrice with respect to $\vect{\theta}$, and the derivatives can be taken under the integral sign. 
\item \label{hp:a4} For each $i=1,\ldots,n$, the matrices $\mat{J}^{(i)}$ are positive definite and
  \begin{equation*}
    \lambda_0 = \inf_n [ \mbox{min eigenvalue of } \mat{\Psi}_n] >0
  \end{equation*}
\item \label{hp:a5} There exists a function $M_{jkl}^{(i)}(\vect{y})$ such that
  \begin{equation*}
    |\nabla_{jkl}H_i(\vect{y}, \vect{\theta})| \le M_{jkl}^{(i)}(\vect{y}) \qquad \forall \vect{\theta} \in \Theta, \mbox{  } \forall i
  \end{equation*}  
  where
  \begin{equation*}
    \frac{1}{n} \sum_{i=1}^n \mathrm{E}_{g_i} [ M_{jkl}^{(i)}(\vect{Y})] = \mathcal{O}(1) \qquad \forall j,k,l
  \end{equation*}
\item \label{hp:a6} For all $j,k$, we have
  \begin{align}
  \label{eqn:a6-1}  
  \lim_{N \rightarrow \infty} \sup_{n>1} \bigg\{ \frac{1}{n}&\sum_{i=1}^n\mathrm{E}_{g_i}[|\nabla_jH_i(\vect{Y}, \vect{\theta})|I(|\nabla_jH_i(\vect{Y}, \vect{\theta})|>N)]\bigg\}=0 \\
  \lim_{N \rightarrow \infty} \sup_{n>1} \bigg\{ \frac{1}{n}&\sum_{i=1}^n\mathrm{E}_{g_i}[|\nabla_{jk}H_i(\vect{Y}, \vect{\theta})-\mathrm{E}_{g_i}(\nabla_{jk}H_i(\vect{Y}, \vect{\theta}))| \nonumber\\
    &\times I(|\nabla_{jk}H_i(\vect{Y}, \vect{\theta})-\mathrm{E}_{g_i}(\nabla_{jk}H_i(\vect{Y}, \vect{\theta}))|>N)] \bigg\}=0
  \label{eqn:a6-2}
  \end{align}
where $I(B)$ denotes the indicator variable of the event $B$.
\item \label{hp:a7} For all $\epsilon > 0$, we have
  \begin{equation}
    \label{eqn:a7}
    \lim_{n \rightarrow \infty} \bigg\{ \frac{1}{n}\sum_{i=1}^n\mathrm{E}_{g_i}\left[||\mat{\Omega}_n^{-1/2}\nabla H_i(\vect{Y}, \vect{\theta})||^2I(||\mat{\Omega}_n^{-1/2}\nabla H_i(\vect{Y}, \vect{\theta})|| > \epsilon\sqrt{n})\right] \bigg\} = 0.
  \end{equation}
\end{enumerate}

\citet{Ghosh2013} proved that, under assumptions \ref{hp:a1}-\ref{hp:a7}, 
the MDPDE $\hat{\vect{\theta}}$ is consistent for $\vect{\theta}$ and asymptotically normal. 
In particular, the asymptotic distribution of $\mat{\Omega}_n^{-1/2}\mat{\Psi}_n[\sqrt{n}(\vect{\theta}_n - \vect{\theta}^g)]$ is 
$p$-dimensional normal with (vector) mean $0$ and covariance matrix $\mat{I}_p$, the $p$-dimensional identity matrix.

\section{Proof of Theorem 1}
\label{sm:sec:proof-lemma}

First let us note that, under the setup of the linear mixed models introduced in Section 3, 
the matrices $\mat{\Omega}_n$ and $\mat{\Psi}_n$ defined in Section \ref{sm:sec:asymptotic} simplifies to the forms 
given in Subsection 3.1 of the main paper and involved in Theorem 1.
Also, the LMM setup is a special case of the general setup of independent non-homogeneous observations and
we assumed that the true data generating density belongs to the model family. 
Then, the proof of our Theorem 1 is immediate from the results stated by \citet{Ghosh2013},
provided we can show that the general Assumptions \ref{hp:a1}-\ref{hp:a7} are satisfied  
under the our assumed Conditions (MM1)--(MM4) for the special case of LMMs.

Now, the assumption that the true data generating density belongs to the model family, 
together with that the model density is normal with mean $\mat{X}_i\vect{\beta}$ and variance matrix $\mat{V_i}$, 
ensures that Assumptions \ref{hp:a1}-\ref{hp:a3} are directly satisfied. 
Also, Assumption \ref{hp:a4} follows from Condition (MM1). 

Next, we prove Equation (\ref{eqn:a6-1}) of Assumption \ref{hp:a6}. 
For any $j = 1,\ldots,p$, the $j$-th partial derivative with respect to $\vect{\beta}$ is given by
\begin{equation*}
  \nabla_jH_i(\vect{Y}_i,\vect{\theta}) = -\alpha \left (1+\frac{1}{\alpha} \right )\eta_{i\alpha}e^{-\frac{\alpha}{2}(\vect{Y}_i - \mat{X}_i\vect{\beta})^\top \mat{V}_i^{-1}(\vect{Y}_i - \mat{X}_i\vect{\beta})}\mat{X}_{ij}^\top \mat{V}_i^{-1}(\vect{Y}_i - \mat{X}_i\vect{\beta}).
\end{equation*}
Then, considering $\vect{Z}_i = \mat{V}_i^{-\frac{1}{2}}(\vect{Y}_i - \mat{X}_i\vect{\beta})$, 
\begin{align*}
&  \frac{(1+\alpha)}{n} \sum_{i=1}^n \mathrm{E}_{g_i}  \left[|\eta_{i\alpha}e^{-\frac{\alpha}{2}(\vect{Y}_i - \mat{X}_i\vect{\beta})^\top \mat{V}_i^{-1}(\vect{Y}_i - \mat{X}_i\vect{\beta})}\mat{X}_{ij}^\top \mat{V}_i^{-1}(\vect{Y}_i - \mat{X}_i\vect{\beta})| \times \right.\\
    & ~~~~~~~~~~~~~~~~~~~~~~ \left.
    I(|\eta_{i\alpha}e^{-\frac{\alpha}{2}(\vect{Y}_i - \mat{X}_i\vect{\beta})^\top \mat{V}_i^{-1}(\vect{Y}_i - \mat{X}_i\vect{\beta})}\mat{X}_{ij}^\top \mat{V}_i^{-1}(\vect{Y}_i - \mat{X}_i\vect{\beta})|> \frac{N}{1+\alpha})
    \right] 
    \\
  \le &  \frac{(1+\alpha)}{n} \sum_{i=1}^n  \eta_{i\alpha}|\mat{X}_{ij}^\top \mat{V}_i^{-\frac{1}{2}}| \mathrm{E}_{g_i}  \left[e^{-\frac{\alpha}{2} \vect{Z}_i^\top \vect{Z}_i} |\vect{Z}_i| \times \right.\\
  & \qquad  \left. I \left (e^{-\frac{\alpha}{2}\vect{Z}_i^\top \vect{Z}_i} |\vect{Z}_i|> \frac{N}{(1+\alpha) (\sup_{n > 1} \eta_{i\alpha}) (\sup_{n > 1} \max_{1 \le i \le n} |\mat{X}_{ij}^\top \mat{V}_i^{-\frac{1}{2}}|)}\right ) \right] \\
  =  & E_1\left [e^{-\frac{\alpha}{2} \vect{Z}_1^\top \vect{Z}_1} |\vect{Z}_1| I \left (e^{-\frac{\alpha}{2}\vect{Z}_1^\top \vect{Z}_1} |\vect{Z}_1|> \frac{N}{(1+\alpha) (\sup_{n > 1} \eta_{i\alpha}) (\sup_{n > 1} \max_{1 \le i \le n} |\mat{X}_{ij}^\top \mat{V}_i^{-\frac{1}{2}}|)}\right ) \right ] \times \\
  & \qquad  \left ( \frac{1+\alpha}{n} \sum_{i=1}^n \eta_{i\alpha}|\mat{X}_{ij}^\top \mat{V}_i^{-\frac{1}{2}}|\right )
\end{align*}
Since the $\sup_{n > 1} \max_{1 \le i \le n} |\mat{X}_{ij}^\top \mat{V}_i^{-\frac{1}{2}}| = \mathcal{O}(1)$ by Assumption (MM2) and $\sup_{n > 1} \eta_{i\alpha}$ is bounded thanks to the boundness of $|\mat{V}_i|$ in (MM3), by the Dominated Convergence Theorem (DCT) we have
\begin{equation*}
  \lim_{N \rightarrow \infty} E_1\left [e^{-\frac{\alpha}{2} \vect{Z}_1^\top \vect{Z}_1} |\vect{Z}_1| I \left (e^{-\frac{\alpha}{2}\vect{Z}_1^\top \vect{Z}_1} |\vect{Z}_1|> \frac{N}{(1+\alpha) (\sup_{n > 1} \eta_{i\alpha}) (\sup_{n > 1} \max_{1 \le i \le n} |\mat{X}_{ij}^\top \mat{V}_i^{-\frac{1}{2}}|)}\right ) \right ] = 0.
\end{equation*}
Also
\begin{equation*}
   \sup_{n > 1} \left ( \frac{1}{n} \sum_{i=1}^n \eta_{i\alpha}|\mat{X}_{ij}^\top \mat{V}_i^{-\frac{1}{2}}|\right ) \le  (\sup_{n > 1} \eta_{i\alpha}) (\sup_{n > 1} \max_{1 \le i \le n} |\mat{X}_{ij}^\top \mat{V}_i^{-\frac{1}{2}}|) = \mathcal{O}(1)
\end{equation*}
and this follows for all $j = 1, \ldots, p$. On the other hand, consider the partial derivative with respect to $\sigma^2_j$, $j=0,1,\ldots,r$, given in Equation (7). Hence, denoting with $\vect{Z}_i = \mat{V}_i^{-\frac{1}{2}}(\vect{Y}_i - \mat{X}_i\vect{\beta})$, we have

\begin{align*}
  & \frac{1}{n} \sum_{i=1}^n \mathrm{E}_{g_i} \left [ |\nabla_{\sigma^2}H_i(\vect{Y}_i,\vect{\theta})|\times I(|\nabla_{\sigma^2_j}H_i(\vect{Y}_i,\vect{\theta})|>N) \right ] \\
  \le & \frac{\alpha}{2n}\sum_{i=1}^n \frac{\eta_{i\alpha}\mathrm{Tr}(\mat{V}_i^{-1}\mat{U}_{ij})}{(1+\alpha)^{\frac{n_i}{2}}}\mathrm{E}_{g_i}[I(|\nabla_{\sigma^2_j}H_i(\vect{Y}_i,\vect{\theta})|>N)] + \\
  & + \frac{1+\alpha}{2n}\sum_{i=1}^n \eta_{i\alpha}\mathrm{Tr}(\mat{V}_i^{-1}\mat{U}_{ij})\mathrm{E}_{g_i}[e^{-\frac{\alpha}{2} \vect{Z}_i^\top \vect{Z}_i}I(|\nabla_{\sigma^2_j}H_i(\vect{Y}_i,\vect{\theta})|>N)] + \\
  & +  \frac{1+\alpha}{2n}\sum_{i=1}^n \eta_{i\alpha}\mathrm{E}_{g_i}[|\vect{Z}_i^\top \mat{V}_i^{-1}\mat{U}_{ij}\mat{V}_i^{-1} \vect{Z}_i| e^{-\frac{\alpha}{2} \vect{Z}_i^\top \mat{V}_i^{-1} \vect{Z}_i}I(|\nabla_{\sigma^2_j}H_i(\vect{Y}_i,\vect{\theta})|>N)] \\
  = & \left ( \frac{\alpha}{2n}\sum_{i=1}^n \frac{\eta_{i\alpha}\mathrm{Tr}(\mat{V}_i^{-1}\mat{U}_{ij})}{(1+\alpha)^{\frac{n_i}{2}}} \right ) \mathrm{E}_1[I(|\nabla_{\sigma^2_j}H_i(\vect{Y}_i,\vect{\theta})|>N)] + \\
  & + \left ( \frac{1+\alpha}{2n}\sum_{i=1}^n \eta_{i\alpha}\mathrm{Tr}(\mat{V}_i^{-1}\mat{U}_{ij}) \right ) \mathrm{E}_1[e^{-\frac{\alpha}{2} \vect{Z}_1^\top \vect{Z}_1}I(|\nabla_{\sigma^2_j}H_i(\vect{Y}_i,\vect{\theta})|>N)] + \\
  & +  \frac{1+\alpha}{2n}\sum_{i=1}^n \eta_{i\alpha}\mathrm{E}_{g_i}[|\vect{Z}_i^\top \mat{V}_i^{-1}\mat{U}_{ij}\mat{V}_i^{-1} \vect{Z}_i| e^{-\frac{\alpha}{2} \vect{Z}_i^\top \mat{V}_i^{-1}\vect{Z}_i}I(|\nabla_{\sigma^2_j}H_i(\vect{Y}_i,\vect{\theta})|>N)]
\end{align*}
where in the last term $\vect{Z}_i = (\vect{Y}_i - \mat{X}_i\vect{\beta})$. Note that
$$
\mathrm{E}_{g_i}[\vect{Z}_i^\top \mat{V}_i^{-1}\mat{U}_{ij}\mat{V}_i^{-1} \vect{Z}_i e^{-\frac{\alpha}{2} \vect{Z}_i^\top \mat{V}_i^{-1} \vect{Z}_i}] = \frac{\mathrm{Tr}(\mat{V}_i^{-1}\mat{U}_{ij})}{(1+\alpha)^{1+\frac{n_i}{2}}},
$$
hence, by Equation (11), 
$$
\mathrm{E}_{g_i}\left[|\vect{Z}_i^\top \mat{V}_i^{-1}\mat{U}_{ij}\mat{V}_i^{-1} \vect{Z}_i| e^{-\frac{\alpha}{2} \vect{Z}_i^\top \mat{V}_i^{-1} \vect{Z}_i}I(|\nabla_{\sigma^2}H_i(\vect{Y}_i,\vect{\theta})|>N)\right] \rightarrow 0, ~~~\mbox{ as } N \rightarrow \infty.
$$ 
The expectation in the first two terms goes to zero as $N \rightarrow \infty$ by the DCT, as before, and the sums are bounded by Equation (11) in condition (MM3). This holds for all $j = 0,\ldots,r$. 

Finally, Assumptions \ref{hp:a5}, \ref{hp:a7} and Equation (\ref{eqn:a6-2}) similarly hold using the Equations (10) and (12), (13), (9) and (11), respectively.

\section{Sensitivity measures}
\label{sm:sec:sensitivity}

\subsection*{Gross Error Sensitivity}

Recall the definition of the gross error sensitivity given in \citet{Ghosh2013}. Considering the influence function of the functional $T_\alpha^{\vect{\beta}}$, we have that
\begin{align*}
  \gamma^u_{i_0}(T_\alpha^{\vect{\beta}},G_1,\ldots,G_n) = & \sup_{\vect{t}} \left \{ \norm{({\mat{X}^\prime}^\top \mat{X}^\prime)^{-1} \mat{X}_{i_0}^\top \mat{V}_{i_0}^{-1}(\vect{t}_{i_0} - \mat{X}_{i_0}\vect{\beta})f_{i_0}(\vect{t}_{i_0};\vect{\theta})^\alpha} \right \}  \\
  = & \frac{\sup_{\vect{Z}} \left \{ \norm{({\mat{X}^\prime}^\top \mat{X}^\prime)^{-1} \mat{X}_{i_0}^\top \mat{V}_{i_0}^{-1/2}\vect{Z}} e^{-\frac{\vect{Z}^\top \vect{Z}}{2}} \right \} }{\sqrt{\alpha}(2\pi)^{\frac{n_i \alpha}{2}} |\mat{V}_{i_0}|^{\frac{\alpha}{2}}},
\end{align*}
where $\vect{Z} = \sqrt{\alpha} \mat{V}_{i_0}^{-1/2} (\vect{t}- \mat{X}_{i_0}\vect{\beta})$. Denoting $\mat{A} = ({\mat{X}^\prime}^\top \mat{X}^\prime)^{-1} \mat{X}_{i_0}^\top \mat{V}_{i_0}^{-1/2}$, we have to find the sup of the function $\norm{\mat{A}\vect{Z}} e^{\frac{\vect{Z}^\top \vect{Z}}{2}}$ with respect to $\vect{Z}$. Then, we compute the derivative with respect to $\vect{Z}$ obtaining
\begin{equation*}
  \frac{\partial ((\vect{Z}^\top \mat{A}^\top \mat{A} \vect{Z})^{1/2}e^{-\frac{\vect{Z}^\top \vect{Z}}{2}})}{\partial \vect{Z}} = \frac{\mat{A}^\top \mat{A} \vect{Z} e^{-\frac{\vect{Z}^\top \vect{Z}}{2}}}{(\vect{Z}^\top \mat{A}^\top \mat{A} \vect{Z})^{1/2}} - (\vect{Z}^\top \mat{A}^\top \mat{A} \vect{Z})^{1/2}e^{-\frac{\vect{Z}^\top \vect{Z}}{2}} \vect{Z} = 0.
\end{equation*}
Finally, multiplying the above equation by $\vect{Z}^\top$ we have
\begin{align*}
  & (\vect{Z}^\top \mat{A}^\top \mat{A} \vect{Z})^{1/2}e^{-\frac{\vect{Z}^\top \vect{Z}}{2}} - (\vect{Z}^\top \mat{A}^\top \mat{A} \vect{Z})^{1/2}e^{-\frac{\vect{Z}^\top \vect{Z}}{2}}\vect{Z}^\top \vect{Z} = 0 \\
  & (\vect{Z}^\top \mat{A}^\top \mat{A} \vect{Z})^{1/2}e^{-\frac{\vect{Z}^\top \vect{Z}}{2}} \left[ 1 - \vect{Z}^\top \vect{Z} \right] = 0. 
\end{align*}
A solution is given by $\vect{Z}$ such that $\vect{Z}^\top \vect{Z} = 1 $, that is $\vect{Z} = \frac{\vect{k}}{\norm{\vect{k}}}$ with $\vect{k} \in \mathbb{R}^{n_{i_0}}$. Hence
\begin{align*}
  \sup_{\vect{Z}} \left \{  \norm{\mat{A}\vect{Z}}e^{-\frac{\vect{Z}^\top \vect{Z}}{2}} \right \} = & e^{-1/2}\sup_{\vect{k}} \left\{ \frac{\norm{\mat{A}\vect{k}}}{\norm{\vect{k}}}\right\} \\
  = & e^{-1/2}\sup_{\vect{k}} \left\{ \frac{\vect{k}^\top \mat{A}^\top \mat{A} \vect{k}}{\vect{k}^\top \vect{k}}\right\}^{\frac{1}{2}}.
\end{align*}
We know that $\sup_{\vect{z}} \left\{ \frac{\vect{z}^\top \mat{A} \vect{z}}{\vect{z}^\top \vect{z}}\right\} = \lambda_{max}(\mat{A})$, where $\lambda_{max}(\mat{A})$ is the maximum eigenvalue of the matrix $\mat{A}$. Using this general results in our case , we obtain
\begin{equation*}
  \gamma_{i_0}^u(T_\alpha^{\vect{\beta}},G_1,\ldots,G_n) = \frac{(\lambda_{max}(({\mat{X}^\prime}^\top \mat{X}^\prime)^{-2}\mat{X}_{i_0}^\top \mat{V}_{i_0}^{-1} \mat{X}_{i_0}))^{\frac{1}{2}}}{\sqrt{e}\sqrt{\alpha}(2\pi)^{\frac{n_i \alpha}{2}} |\mat{V}_{i_0}|^{\frac{\alpha}{2}}}
\end{equation*}

\noindent\textbf{Special Case:} 
When $n_i = p$, for all $i=1, \ldots, n$, the matrix ${\mat{X}^\prime}^\top \mat{X}^\prime$ can be rewritten as
\begin{align*}
  {\mat{X}^\prime}^\top \mat{X}^\prime = & \sum_{i=1}^n \frac{\mat{X}_i^\top \mat{V}^{-1}\mat{X}_i}{(1+\alpha)^{\frac{p}{2}+1}(2\pi)^{\frac{p\alpha}{2}}|\mat{V}|^{\frac{\alpha}{2}}} \\
  = & \frac{1}{(1+\alpha)^{\frac{p}{2}+1}(2\pi)^{\frac{p\alpha}{2}}|\mat{V}|^{\frac{\alpha}{2}}} \sum_{i=1}^n \mat{X}_i^\top \mat{V}^{-1}\mat{X}_i 
\end{align*}
Substituting this simpler form in the formula of the gross error sensitivity we get Equation (21) of the main paper.

\subsection*{Self-Standardized Sensitivity}

Starting from the definition of the self-standardized sensitivity, we have that
\begin{equation*}
  \gamma_{i_0}^s (T_\alpha^{\vect{\beta}}, G_1, \ldots, G_n) = \frac{1}{n} \sup_{\vect{t}} \left\{ (\vect{t}-\mat{X}_{i_0}\vect{\beta})^\top \mat{V}_{i_0}^{-1} \mat{X}_{i_0}({\mat{X}^\ast}^\top \mat{X}^\ast)^{-1}\mat{X}_{i_0}^\top \mat{V}_{i_0}^{-1}(\vect{t}-\mat{X}_{i_0}\vect{\beta})f_i^{2\alpha}(\vect{t},\vect{\theta})\right\}^{\frac{1}{2}} .
\end{equation*}
Let $\vect{Z}=\left(\frac{\mat{V}_{i_0}}{2\alpha}\right)^{-1/2}(\vect{t}-\mat{X}_{i_0}\vect{\beta})$, then the above equation has the the form
\begin{align*}
  \gamma_{i_0}^s (T_\alpha^{\vect{\beta}}, G_1, \ldots, G_n)= & \frac{1}{n \sqrt{2\alpha}(2\pi)^{\frac{n_i \alpha}{2}} |\mat{V}_{i_0}|^{\frac{\alpha}{2}}} \sup_{\vect{Z}} \left\{ \vect{Z}^\top \mat{V}_{i_0}^{-1/2} \mat{X}_{i_0}({\mat{X}^\ast}^\top \mat{X}^\ast)^{-1}\mat{X}_{i_0}^\top \mat{V}_{i_0}^{-1/2} \vect{Z} e^{-\frac{\vect{Z}^\top \vect{Z}}{2}}\right\}^{\frac{1}{2}} \\
    = & \frac{1}{n \sqrt{2\alpha}(2\pi)^{\frac{n_i \alpha}{2}} |\mat{V}_{i_0}|^{\frac{\alpha}{2}}} \sup_{\vect{Z}} \left\{ \vect{Z}^\top \mat{A} \vect{Z} e^{-\frac{\vect{Z}^\top \vect{Z}}{2}}\right\}^{\frac{1}{2}},
\end{align*}
where $\mat{A}= \mat{V}_{i_0}^{-1/2} \mat{X}_{i_0}({\mat{X}^\ast}^\top \mat{X}^\ast)^{-1}\mat{X}_{i_0}^\top \mat{V}_{i_0}^{-1/2}$. In order to find this sup, we compute the derivative with respect to $\vect{Z}$, obtaining
\begin{align*}
  \frac{\partial(\vect{Z}^\top \mat{A} \vect{Z} e^{-\frac{\vect{Z}^\top \vect{Z}}{2}})}{\partial \vect{Z}} = & \vect{Z}^\top \left [2\mat{A}\vect{Z} e^{-\frac{\vect{Z}^\top \vect{Z}}{2}} - \vect{Z}^\top \mat{A} \vect{Z} e^{-\frac{\vect{Z}^\top \vect{Z}}{2}} \vect{Z} \right] \\
  = & \vect{Z}^\top \mat{A} \vect{Z} e^{-\frac{\vect{Z}^\top \vect{Z}}{2}} [2 - \vect{Z}^\top \vect{Z}] = 0.
\end{align*}
A solution is given by $\vect{Z}$ such that $\vect{Z}^\top \vect{Z} = 2$, then $\vect{Z} = \frac{\sqrt{2}\vect{k}}{\norm{\vect{k}}}$, with $\vect{k} \in \mathbb{R}^{n_{i_0}}$. Note that we multiplied by $\vect{Z}^\top$ in the same way done for the gross error sensitivity. Hence
\begin{align*}
  \gamma_{i_0}^s (T_\alpha^{\vect{\beta}}, G_1, \ldots, G_n)= & \frac{\sqrt{2}}{n \sqrt{2\alpha}(2\pi)^{\frac{n_i \alpha}{2}} |\mat{V}_{i_0}|^{\frac{\alpha}{2}}} \sup_{\vect{k}} \left\{ \frac{\vect{k}^\top \mat{A}\vect{k}}{\norm{\vect{k}}^2}\right\}^{\frac{1}{2}} \\
  = & \frac{\left[\lambda_{max}\left(({\mat{X}^\ast}^\top \mat{X}^\ast)^{-1}\mat{X}_{i_0}^\top \mat{V}_{i_0}^{-1}\mat{X}_{i_0} \right)\right]^{1/2} }{n \sqrt{\alpha} (2\pi)^{\frac{n_i\alpha}{2}}|\mat{V}_i|^{\frac{\alpha}{2}} e^{1/2}},
\end{align*}
which corresponds to equation (19). 

\noindent\textbf{Special Case:} 
When $n_i = p$, for all $i=1, \ldots, n$, the matrix ${\mat{X}^\ast}^\top \mat{X}^\ast$ can be rewritten as
\begin{align*}
  {\mat{X}^\ast}^\top \mat{X}^\ast = & \sum_{i=1}^n \frac{\mat{X}_i^\top \mat{V}^{-1}\mat{X}_i}{(1+\alpha)^{\frac{p}{2}+1}(2\pi)^{p\alpha}|\mat{V}|^{\alpha}} \\
  = & \frac{1}{(1+\alpha)^{\frac{p}{2}+1}(2\pi)^{p\alpha}|\mat{V}|^{\alpha}} \sum_{i=1}^n \mat{X}_i^\top \mat{V}^{-1}\mat{X}_i
\end{align*}
Substituting this simpler form in the formula of the self-standardized sensitivity we get equation (22).

\section{Additional Numerical Results }
\label{sm:sec:numerical-study}

In this section we provide the plots of Asymptotic Relative Efficiency and Influence functions,
corresponding to the Example in Section 3.3 of the main paper.

\begin{figure}
\begin{center}
\includegraphics[width=0.45\textwidth]{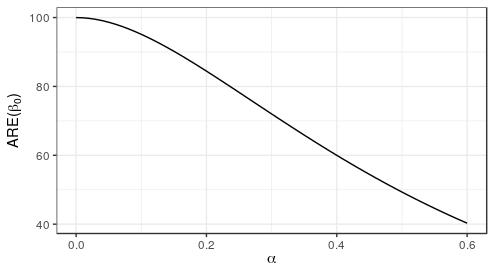}
\includegraphics[width=0.45\textwidth]{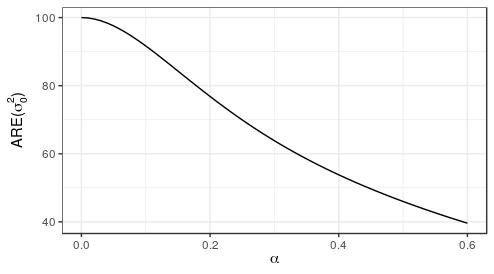}
\includegraphics[width=0.45\textwidth]{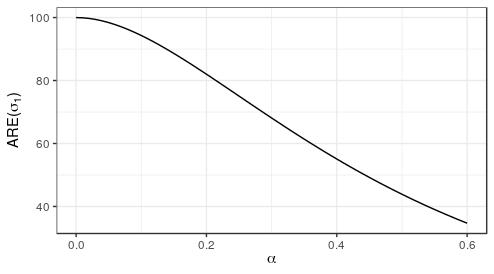}
\end{center}
\caption{Asymptotic Relative Efficiency with respect to $\alpha$ for $\beta_0$, $\sigma_0^2$ and $\sigma^2_1$, respectively.}
\label{fig:are-other}
\end{figure}

\begin{figure}[hbt!]
\begin{center}
\includegraphics[width=0.49\textwidth]{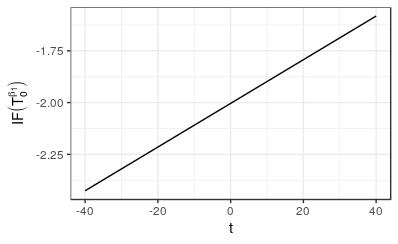}
\includegraphics[width=0.49\textwidth]{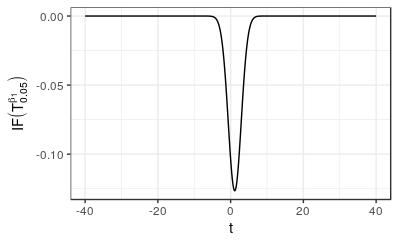}
\includegraphics[width=0.49\textwidth]{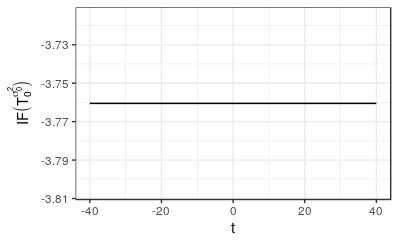}
\includegraphics[width=0.49\textwidth]{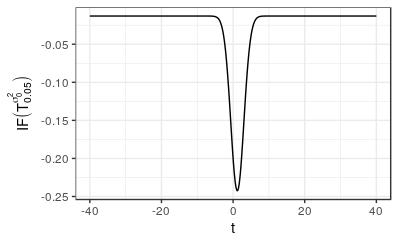}
\includegraphics[width=0.49\textwidth]{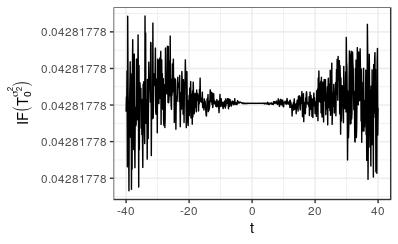}
\includegraphics[width=0.49\textwidth]{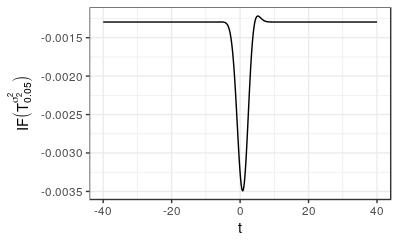}
\end{center}
\caption{Influence function for the estimators $T_\alpha^{\beta_1}$, $T_\alpha^{\sigma_0^2}$ and $T_\alpha^{\sigma_2^2}$ with respect to $\alpha = 0$ (on the left) and $\alpha = 0.05$ (on the right).}
\label{fig:if-other-0-005}
\end{figure}

\begin{figure}[hbt!]
\begin{center}
\includegraphics[width=0.49\textwidth]{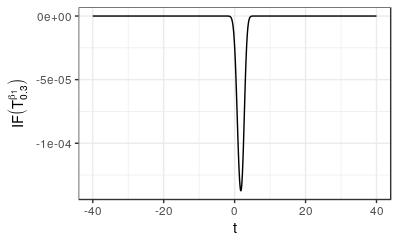}
\includegraphics[width=0.49\textwidth]{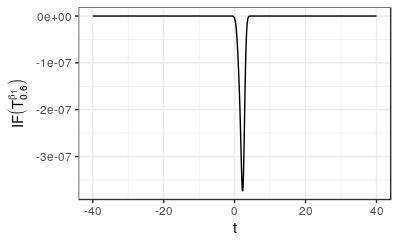}
\includegraphics[width=0.49\textwidth]{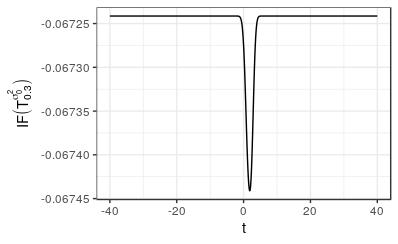}
\includegraphics[width=0.49\textwidth]{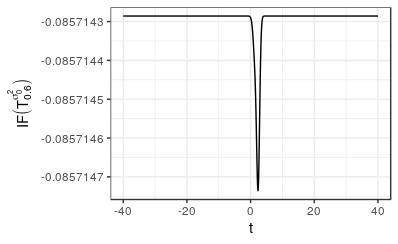}
\includegraphics[width=0.49\textwidth]{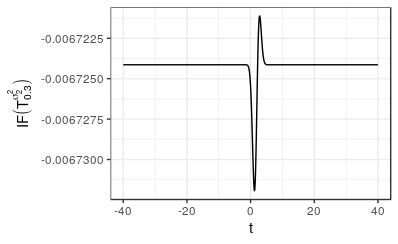}
\includegraphics[width=0.49\textwidth]{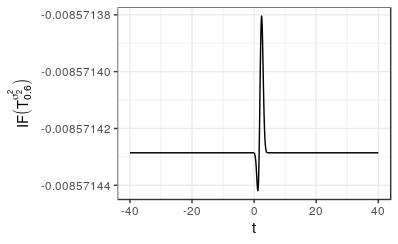}
\end{center}
\caption{Influence function for the estimators $T_\alpha^{\beta_1}$, $T_\alpha^{\sigma_0^2}$ and $T_\alpha^{\sigma_2^2}$ with respect to $\alpha = 0.3$ (on the left) and $\alpha = 0.6$ (on the right).}
\label{fig:if-other-03-06}
\end{figure}

Figure \ref{fig:are-other} shows the Asymptotic Relative Efficiency with respect to $\alpha$ for $\beta_0$, $\sigma_0^2$ and $\sigma^2_1$, respectively.
In Figure \ref{fig:if-other-0-005} we can see the influence function for the estimators $T_\alpha^{\beta_1}$, $T_\alpha^{\sigma_0^2}$ and $T_\alpha^{\sigma_2^2}$ for $\alpha = 0, 0.05$. While Figure \ref{fig:if-other-03-06} presents the same quantities for $\alpha = 0.3, 0.6$.

\section{Additional Results from Monte Carlo experiment}
\label{sm:sec:montecarlo}

Here, we report the complete results of the Monte Carlo experiments with respect to the contamination levels considered. 

\begin{figure}[ht]
\begin{center}
	\subfloat[MSMD performance of the MDPD-estimators of $\vect{\beta}$]{
\includegraphics[width=0.5\textwidth]{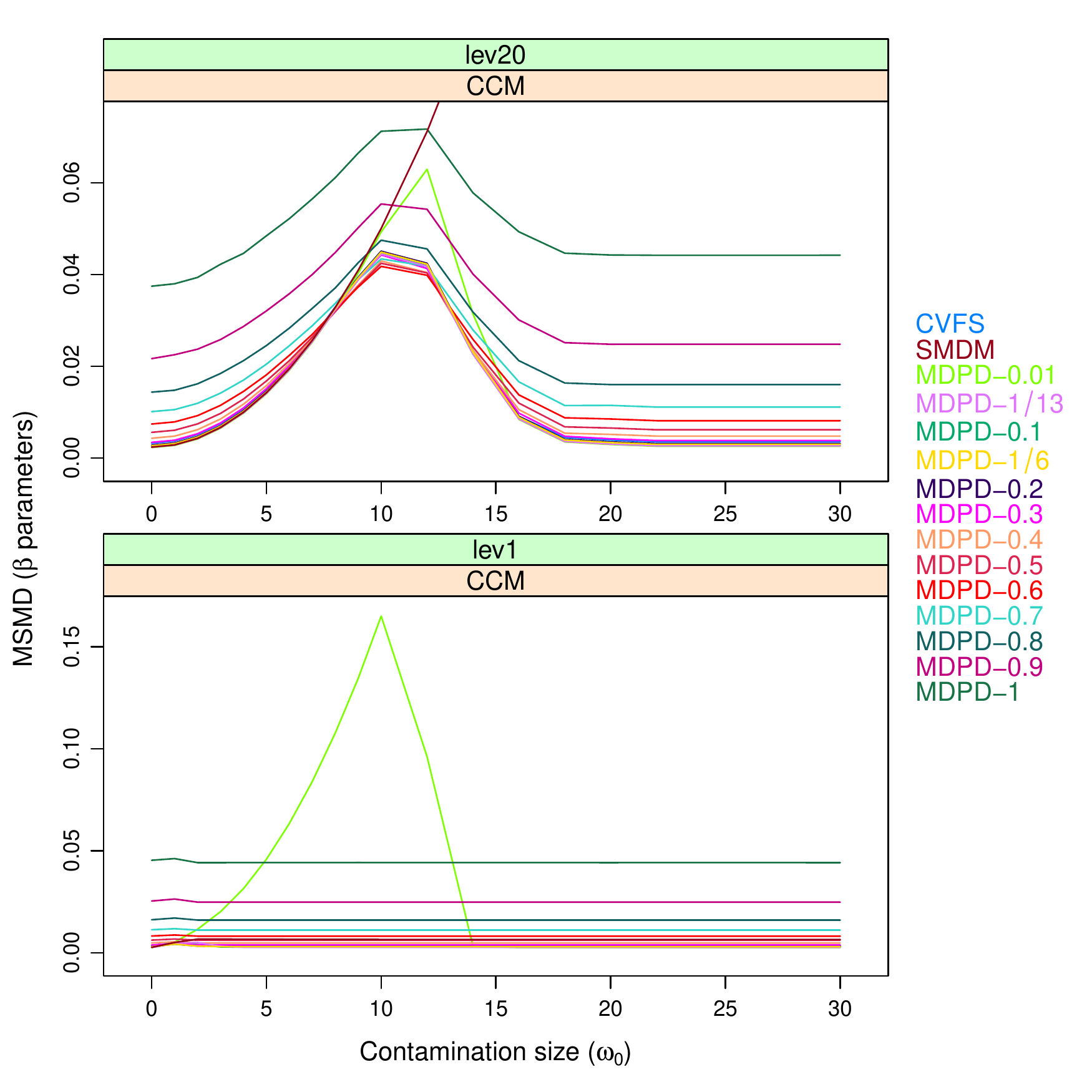}
		\label{fig:msmd-5}} ~
	\subfloat[MKLD performance of the MDPD-estimators of $(\eta,\vect{\gamma})$]{
	\includegraphics[width=0.5\textwidth]{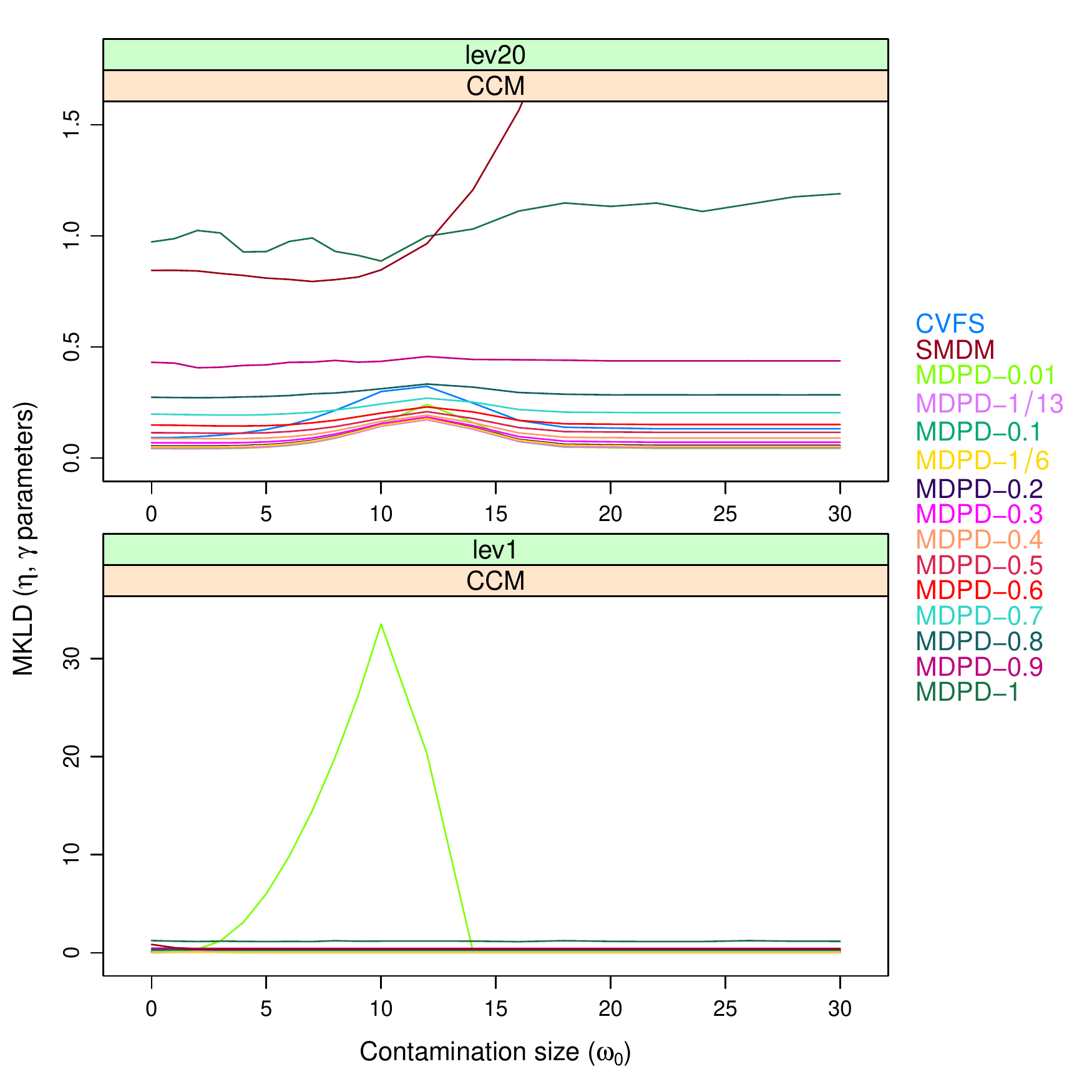}
	\label{fig:mkld-5}} ~
\caption{Performance of the MDPD-estimators of $\vect{\beta}$ and $(\eta,\vect{\gamma})$ 
	considering different values of $\alpha$ (including $\alpha^*=1/13$ and $\bar{\alpha}=1/6$)
	compared to the CVFS- and SMDM-estimators, under $5\%$  outlier contamination.}
\end{center}
\label{fig:sim5}
\end{figure}

The MSMD and MKLD performances of the CVFS-, SMDM-estimators and for MDPDE for different values of $\alpha$ with $5\%$ of contamination are presented in Figure \ref{fig:msmd-5} and \ref{fig:mkld-5}, respectively.

\begin{figure}[htbp]
\begin{center}
	\subfloat[MSMD performance of the MDPD-estimators of $\vect{\beta}$]{
\includegraphics[width=0.5\textwidth]{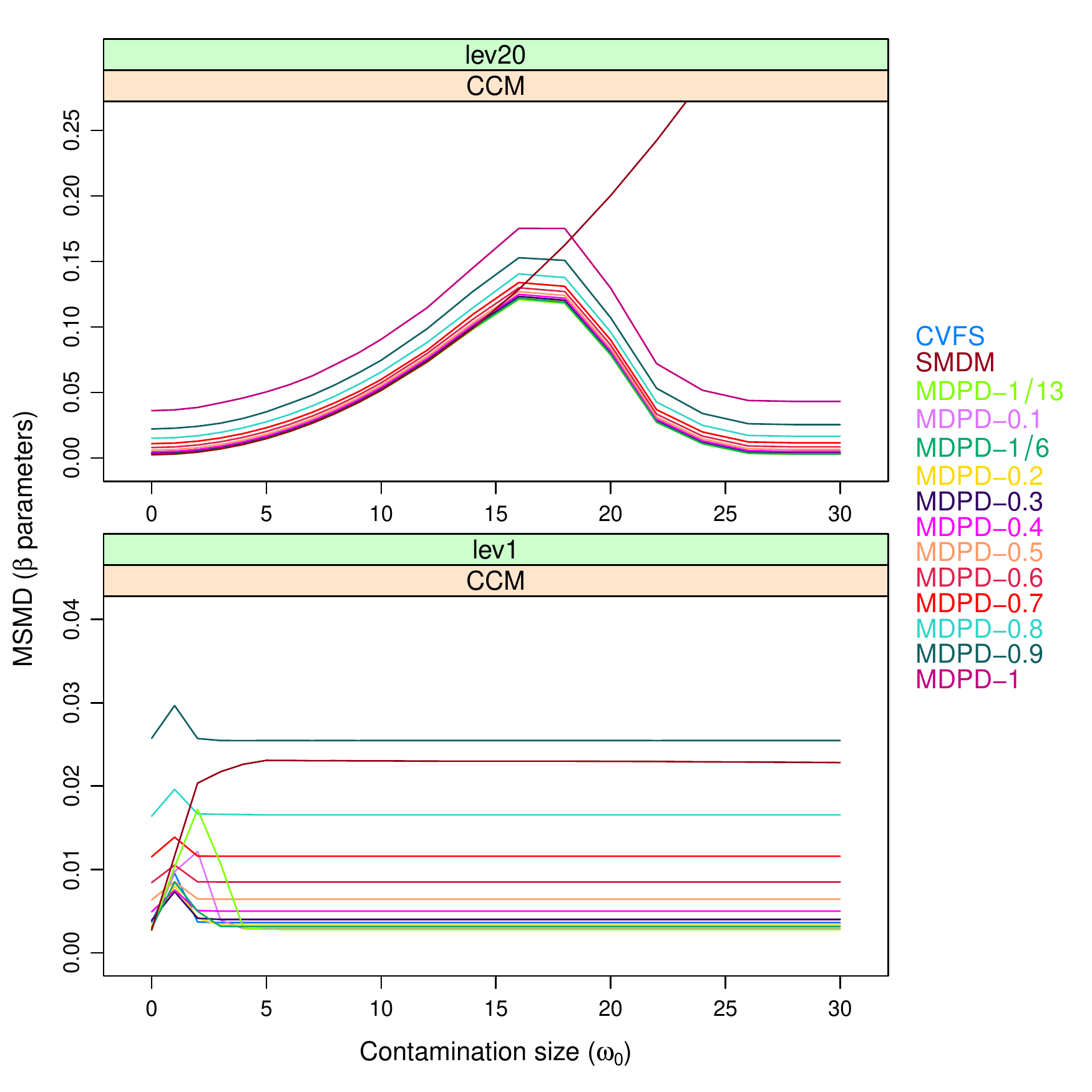}
		\label{fig:msmd-10}} ~
	\subfloat[MKLD performance of the MDPD-estimators of $(\eta,\vect{\gamma})$]{
	\includegraphics[width=0.5\textwidth]{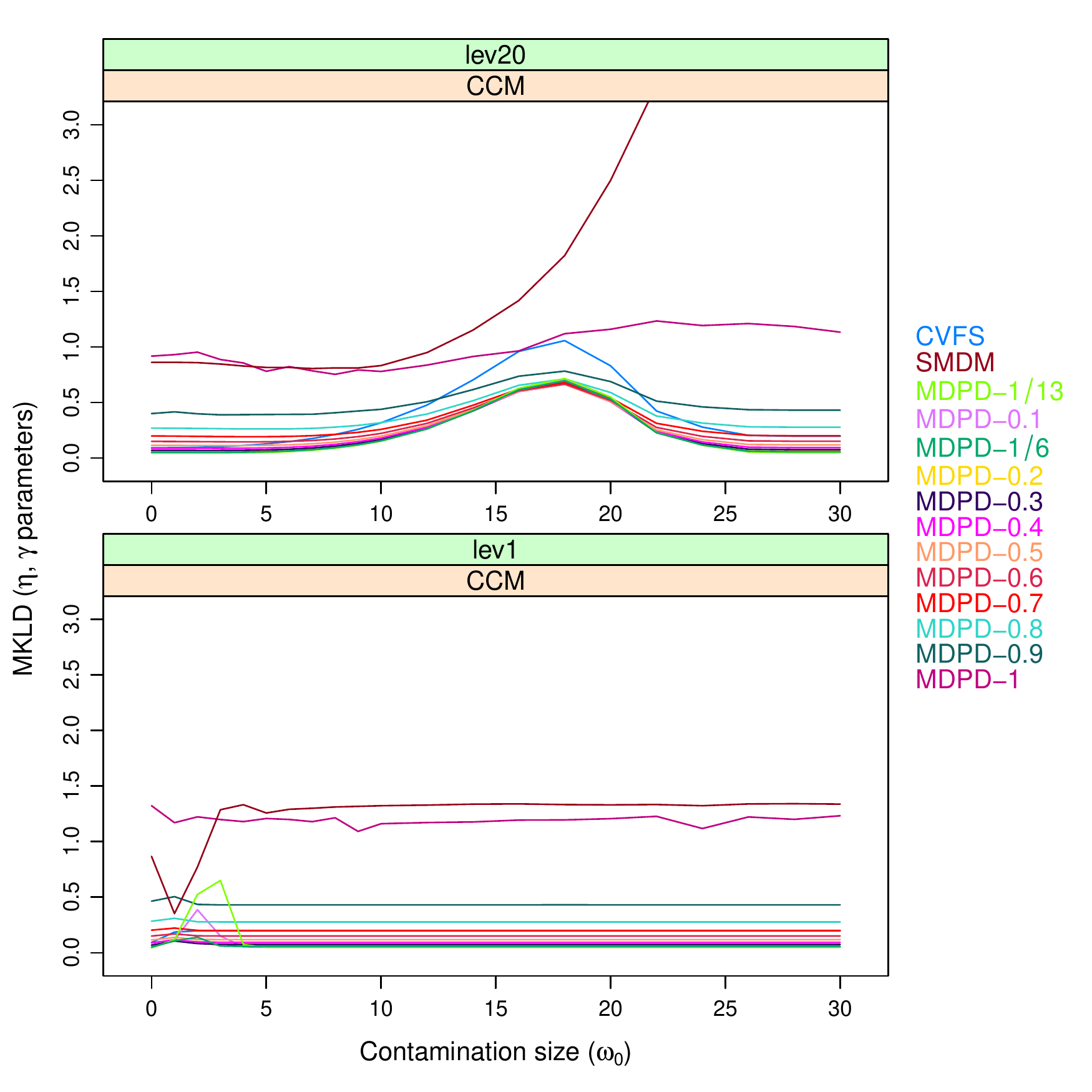}
	\label{fig:mkld-10}} ~
\caption{Performance of the MDPD-estimators of $\vect{\beta}$ and $(\eta,\vect{\gamma})$ 
	considering different values of $\alpha$ (including $\alpha^*=1/13$ and $\bar{\alpha}=1/6$)
	compared to the CVFS- and SMDM-estimators, under $10\%$  outlier contamination.}
\end{center}
\label{fig:sim10}
\end{figure}

Figure \ref{fig:msmd-10} and \ref{fig:mkld-10} display the MSMD and MKLD performances of the CVFS-, SMDM-estimators and for MDPDE for different values of $\alpha$ with $10\%$ of contamination.

\begin{figure}[htbp]
\begin{center}
	\subfloat[MSMD performance of the MDPD-estimators of $\vect{\beta}$]{
\includegraphics[width=0.5\textwidth]{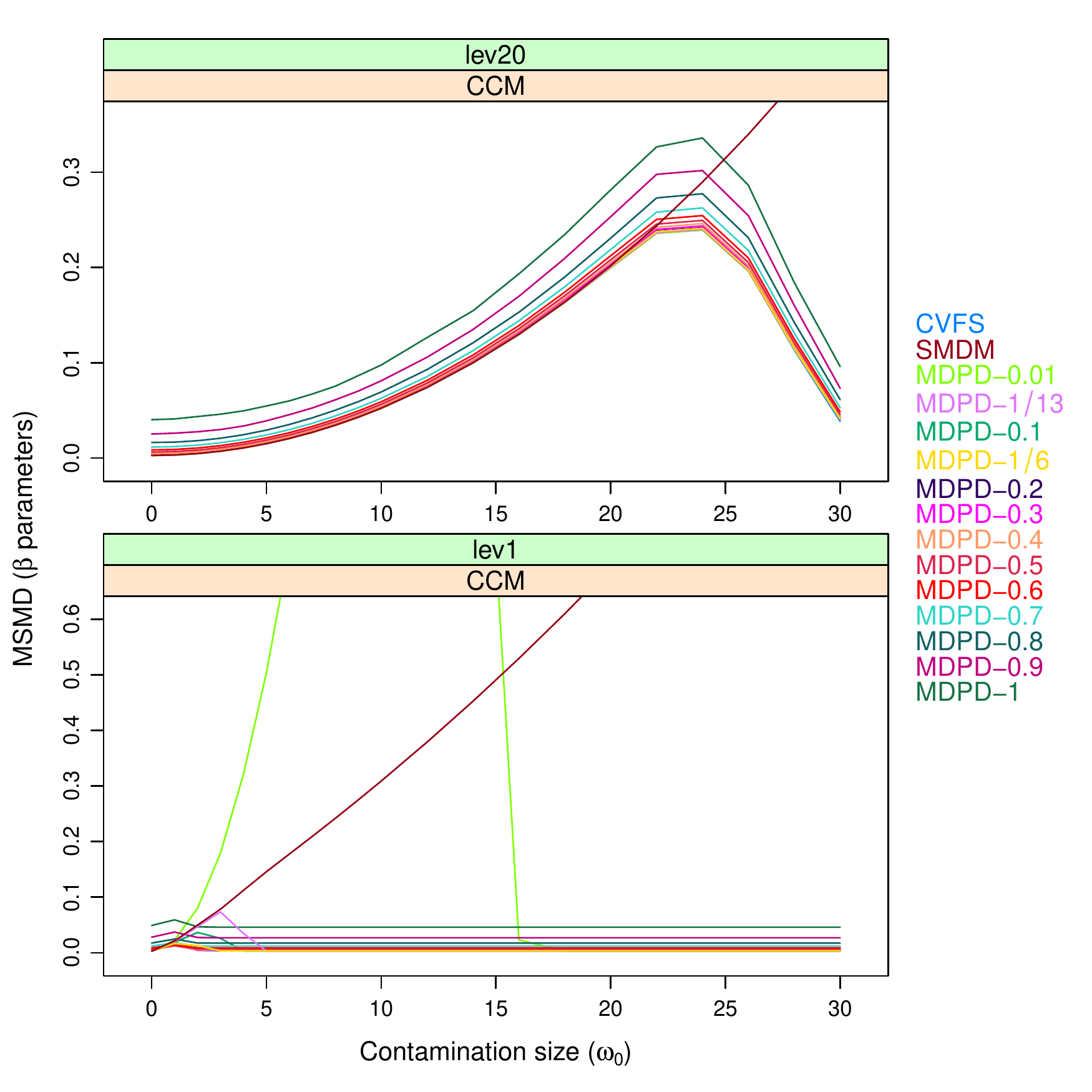}
		\label{fig:msmd-15}} ~
	\subfloat[MKLD performance of the MDPD-estimators of $(\eta,\vect{\gamma})$]{
	\includegraphics[width=0.5\textwidth]{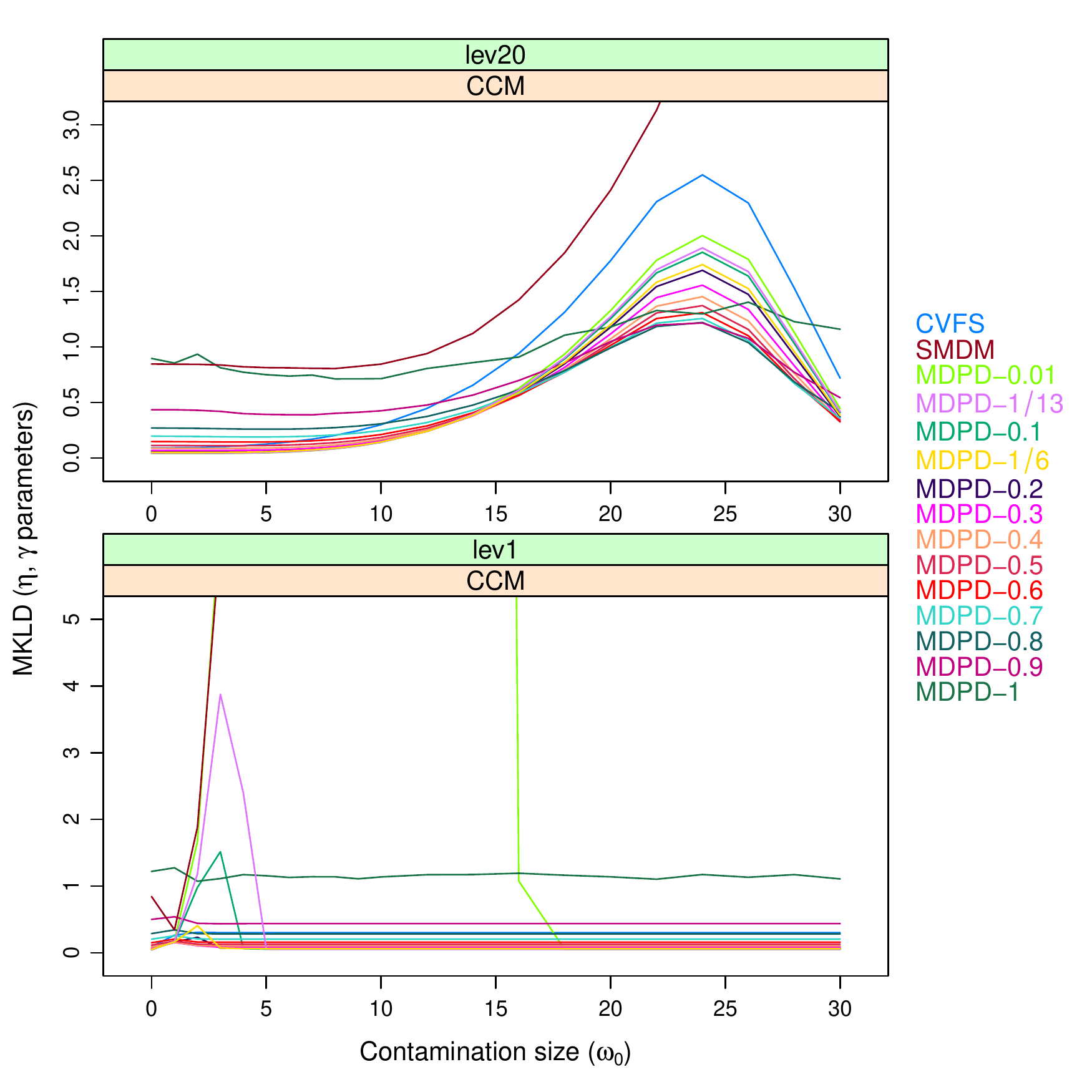}
	\label{fig:mkld-15}} ~
\caption{Performance of the MDPD-estimators of $\vect{\beta}$ and $(\eta,\vect{\gamma})$ 
	considering different values of $\alpha$ (including $\alpha^*=1/13$ and $\bar{\alpha}=1/6$)
	compared to the CVFS- and SMDM-estimators, under $15\%$  outlier contamination.}
\end{center}
\label{fig:sim15}
\end{figure}

While Figures \ref{fig:msmd-15} and \ref{fig:mkld-15} show the MSMD and MKLD performances, respectively, of the CVFS-, SMDM-estimators and for MDPDE for different values of the parameter $\alpha$ in case of $15\%$ of outlier contamination.

\begin{table}[htbp]
\centering
\begin{tabular}{rr |rr|rr}
     \hline
	&&	\multicolumn{2}{c|}{MSMD} & 	\multicolumn{2}{c }{MKLD}\\
		Method & $(\alpha)$ & lev1 & lev20 & lev1 & lev20\\
		\hline
     CVFS & -               &   0.005  &  0.045 &   0.132  &  0.322 \\
     SMDM & -               &   0.007  &  0.444 &   0.843  &  9.347 \\
     MDPDE & 0              &   2.255  &  2.255 &   2.276e22  & 8.198e25 \\
      & 0.01           &   0.165  &  0.063 &   33.537  &  0.241 \\
     & $\alpha^\ast$   &   0.005  &  0.045 &   0.146  &  0.172 \\
     & 0.1            &   0.004  &  0.044 &   0.107  &  0.173 \\
     & $\bar{\alpha}$ &   0.004  &  0.045 &   0.066  &  0.177 \\
     & 0.2            &   0.004  &  0.045 &   0.067  &  0.179 \\
     & 0.3            &   0.005  &  0.044 &   0.077  &  0.184 \\
     & 0.4            &   0.005  &  0.043 &   0.095  &  0.192 \\
     & 0.5            &   0.007  &  0.042 &   0.121  &  0.209 \\
     & 0.6            &   0.009  &  0.042 &   0.157  &  0.230 \\
   \hline
   \end{tabular}
\caption{Maximum values of MSMD and MKLD for the CVFS-, SMDM-estimators and for MDPDE considering different values of $\alpha$ under $5\%$ of outlier contamination.}
\label{tab:max-5}
\end{table}

\begin{table}[htbp]
\centering
\begin{tabular}{rr |rr|rr}
  \hline
		&&	\multicolumn{2}{c|}{MSMD} & 	\multicolumn{2}{c }{MKLD}\\
		Method & $(\alpha)$ & lev1 & lev20 & lev1 & lev20\\
		\hline
     CVFS    & -             &   0.019  &  0.240 &   0.306  &  2.549 \\
     SMDM    & -             &   1.174  &  0.452 &   225.429  &  8.162 \\
     MDPDE & 0              &   20.260 &  20.275 &   1.439e23  & 1.606e22 \\    
     & 0.01           &   2.892  &  0.239 &   164.868  &  2.002 \\
     & $\alpha^\ast$   &   0.240  &  0.240 &   3.878  &  1.891 \\
     & 0.1            &   0.036  &  0.240 &   1.516  &  1.852 \\
     & $\bar{\alpha}$ &   0.241  &  0.241 &   0.405  &  1.741 \\
     & 0.2            &   0.016  &  0.242 &   0.232  &  1.689 \\
     & 0.3            &   0.013  &  0.244 &   0.155  &  1.556 \\
     & 0.4            &   0.012  &  0.246 &   0.159  &  1.452 \\
     & 0.5            &   0.013  &  0.249 &   0.174  &  1.372 \\
     & 0.6            &   0.014  &  0.255 &   0.203  &  1.308 \\
   \hline
   \end{tabular}
\caption{Maximum values of MSMD and MKLD for the CVFS-, SMDM-estimators and for MDPDE considering different values of $\alpha$ under $15\%$ of outlier contamination.}
\label{tab:max-15}
\end{table}

Tables \ref{tab:max-5} and \ref{tab:max-15} report the maximum values of MSMD and MKLD of the CVFS-, SMDM-estimators and for MDPDE considering different values of $\alpha$ in case of $5\%$ and $15\%$ of contamination, respectively.

\section{Results from real-data example}
\label{sm:sec:example}
In this Section, we report the complete results from the study of the real-data example about Extrafoveal Vision Acuity presented in Section 5.

Table \ref{tab:reading-estimates} shows the estimates of model parameters obtained using the \texttt{lmer} estimator, the SMDM- and CVFS-estimators, the MDPDE with different values of $\alpha$. 

\begin{table}[htbp]
	\centering
	\resizebox{\textwidth}{!}{
	\begin{tabular}{rrrrrrrrrr}
		\hline \\
 Method & lmer & SMDM & CVFS & \multicolumn{6}{c}{MDPDE} \\
         $(\alpha)$        & - & - & - & 0.05 & 1/13 & 1/6 & 0.2 & 0.4 & 0.6 \\

		\hline
\multicolumn{10}{l}{$\beta$}\\	
Intercept	&	7.007	&	7.034	&	7.041	&	7.040	&	7.060	&	7.123	&	7.151	&	7.196	&	7.197	\\
EV	&	7.791	&	7.826	&	7.884	&	7.828	&	7.834	&	7.798	&	7.766	&	7.643	&	7.722	\\
H	&	-0.021	&	-0.009	&	-0.001	&	0.000	&	0.002	&	-0.007	&	-0.010	&	-0.048	&	-0.030	\\
T2\_1	&	0.062	&	0.024	&	-0.016	&	0.002	&	-0.019	&	-0.068	&	-0.090	&	-0.165	&	-0.018	\\
T3\_2	&	-0.081	&	-0.066	&	-0.033	&	-0.040	&	-0.030	&	-0.015	&	-0.009	&	0.043	&	0.189	\\
EV*H	&	0.439	&	0.454	&	0.495	&	0.459	&	0.449	&	0.372	&	0.333	&	0.205	&	-0.012	\\
EV*T2\_1	&	0.180	&	0.079	&	0.046	&	0.084	&	0.054	&	0.003	&	-0.015	&	-0.070	&	0.036	\\
EV*T3\_2	&	-0.204	&	-0.139	&	-0.114	&	-0.122	&	-0.109	&	-0.104	&	-0.101	&	-0.036	&	0.090	\\
H*T2\_1	&	0.144	&	0.142	&	0.099	&	0.122	&	0.118	&	0.124	&	0.123	&	0.106	&	0.468	\\
H*T3\_2	&	-0.162	&	-0.122	&	-0.092	&	-0.121	&	-0.111	&	-0.109	&	-0.112	&	-0.134	&	-0.270	\\
EV*H*T2\_1	&	0.261	&	0.243	&	0.153	&	0.215	&	0.195	&	0.148	&	0.121	&	-0.007	&	0.504	\\
EV*H*T3\_2	&	-0.476	&	-0.370	&	-0.319	&	-0.365	&	-0.335	&	-0.276	&	-0.243	&	-0.152	&	-0.396	\\
\hline
\multicolumn{10}{l}{$\sigma^2$}\\
Intercept	&	0.987	&	1.078	&	0.130	&	0.120	&	0.109	&	0.074	&	0.055	&	0.010	&	0.000	\\
EV	&	1.128	&	1.010	&	0.129	&	0.154	&	0.140	&	0.125	&	0.118	&	0.064	&	0.000	\\
H	&	0.613	&	0.618	&	0.071	&	0.061	&	0.063	&	0.071	&	0.075	&	0.088	&	0.000	\\
T2\_1	&	0.296	&	0.000	&	0.010	&	0.012	&	0.012	&	0.012	&	0.012	&	0.011	&	0.000	\\
T3\_2	&	0.452	&	0.000	&	0.005	&	0.019	&	0.016	&	0.011	&	0.011	&	0.013	&	0.000	\\
EV*H	&	0.656	&	0.000	&	0.063	&	0.050	&	0.048	&	0.033	&	0.021	&	0.024	&	0.000	\\
EV*T2\_1	&	0.000	&	0.000	&	0.000	&	0.000	&	0.000	&	0.000	&	0.000	&	0.000	&	0.000	\\
EV*T3\_2	&	1.032	&	0.000	&	0.101	&	0.103	&	0.091	&	0.061	&	0.046	&	0.000	&	0.000	\\
H*T2\_1	&	0.000	&	0.000	&	0.000	&	0.000	&	0.000	&	0.000	&	0.006	&	0.058	&	0.000	\\
H*T3\_2	&	0.439	&	0.000	&	0.000	&	0.008	&	0.003	&	0.000	&	0.000	&	0.000	&	0.000	\\
$\eta_0$	&	0.154	&	0.118	&	0.121	&	0.120	&	0.111	&	0.099	&	0.095	&	0.082	&	0.000	\\
		\hline
	\end{tabular}}
	\caption{Estimates of model parameters obtained using different estimators.}
	\label{tab:reading-estimates}
\end{table}


\begin{table}[htbp]
  \centering
  \resizebox{\textwidth}{!}{
    \begin{tabular}{rrrrrrrrrrr}
  \hline
  & & $m=1$ & $m=2$ & $m=3$ & $m=4$ & $m=5$ & $m=6$ & $m=7$ & $m=8$ & $m=9$ \\ 
  \hline
$\beta$ & Intercept & 7.001 & 6.994 & 6.987 & 6.963 & 6.940 & 6.917 & 6.896 & 6.890 & 6.882 \\ 
       &   EV & 7.801 & 7.815 & 7.830 & 7.783 & 7.736 & 7.691 & 7.649 & 7.662 & 7.676 \\ 
  & H & -0.032 & -0.017 & -0.002 & 0.046 & -0.001 & 0.045 & 0.087 & 0.099 & 0.085 \\ 
  & T2\_1 & 0.046 & 0.068 & 0.068 & 0.139 & 0.070 & 0.070 & 0.007 & 0.026 & 0.048 \\ 
  & T3\_2 & -0.065 & -0.065 & -0.088 & -0.088 & -0.018 & -0.086 & -0.023 & -0.023 & -0.023 \\ 
  & EV*H & 0.460 & 0.431 & 0.401 & 0.497 & 0.404 & 0.495 & 0.579 & 0.553 & 0.583 \\ 
  & EV*T2\_1 & 0.211 & 0.168 & 0.168 & 0.311 & 0.172 & 0.172 & 0.046 & 0.008 & -0.036 \\ 
  & EV*T3\_2 & -0.235 & -0.235 & -0.190 & -0.190 & -0.051 & -0.187 & -0.061 & -0.061 & -0.061 \\ 
  & H*T2\_1 & 0.113 & 0.070 & 0.070 & -0.074 & -0.212 & -0.212 & -0.087 & -0.125 & -0.081 \\ 
  & H*T3\_2 & -0.130 & -0.130 & -0.086 & -0.086 & 0.053 & 0.189 & 0.063 & 0.063 & 0.063 \\ 
  & EV*H*T2\_1 & 0.323 & 0.410 & 0.410 & 0.123 & -0.155 & -0.155 & 0.097 & 0.173 & 0.085 \\ 
  & EV*H*T3\_2 & -0.538 & -0.538 & -0.628 & -0.628 & -0.351 & -0.079 & -0.330 & -0.330 & -0.330 \\ 
  \hline
$\sigma^2$  & Intercept & 0.952 & 0.909 & 0.911 & 0.590 & 0.450 & 0.495 & 0.473 & 0.466 & 0.456 \\ 
  & EV & 0.974 & 0.985 & 1.049 & 0.814 & 0.520 & 0.609 & 0.481 & 0.472 & 0.464 \\ 
  & H & 0.608 & 0.544 & 0.673 & 0.412 & 0.176 & 0.000 & 0.000 & 0.000 & 0.001 \\ 
  & T2\_1 & 0.009 & 0.407 & 0.518 & 0.557 & 0.216 & 0.136 & 0.003 & 0.001 & 0.000 \\ 
  & T3\_2 & 0.412 & 0.307 & 0.633 & 0.176 & 0.000 & 0.176 & 0.004 & 0.000 & 0.000 \\ 
  & EV*H & 0.603 & 0.990 & 1.153 & 0.961 & 0.225 & 0.690 & 0.617 & 0.644 & 0.618 \\ 
  & EV*T2\_1 & 0.037 & 0.136 & 0.476 & 1.272 & 0.625 & 0.132 & 0.000 & 0.020 & 0.013 \\ 
  & EV*T3\_2 & 0.893 & 1.152 & 1.354 & 0.386 & 0.044 & 0.911 & 0.707 & 0.698 & 0.674 \\ 
  & H*T2\_1 & 0.004 & 0.001 & 0.000 & 1.133 & 0.444 & 0.257 & 0.000 & 0.003 & 0.000 \\ 
  & H*T3\_2 & 0.000 & 0.000 & 0.000 & 0.000 & 0.000 & 0.000 & 0.002 & 0.001 & 0.001 \\ 
  & $\eta_0$ & 0.174 & 0.182 & 0.175 & 0.369 & 0.715 & 0.836 & 1.078 & 1.095 & 1.122 \\ 
  \hline
  $p$-value & Intercept & 0.000 & 0.000 & 0.000 & 0.000 & 0.000 & 0.000 & 0.000 & 0.000 & 0.000 \\ 
  & EV & 0.000 & 0.000 & 0.000 & 0.000 & 0.000 & 0.000 & 0.000 & 0.000 & 0.000 \\ 
  & H & 0.571 & 0.749 & 0.970 & 0.508 & 0.993 & 0.594 & 0.362 & 0.300 & 0.383 \\ 
  & T2\_1 & 0.324 & 0.224 & 0.249 & 0.113 & 0.481 & 0.504 & 0.952 & 0.824 & 0.685 \\ 
  & T3\_2 & 0.232 & 0.215 & 0.169 & 0.216 & 0.846 & 0.416 & 0.841 & 0.842 & 0.844 \\ 
  & EV*H & 0.000 & 0.000 & 0.001 & 0.001 & 0.014 & 0.015 & 0.010 & 0.016 & 0.011 \\ 
  & EV*T2\_1 & 0.025 & 0.086 & 0.096 & 0.095 & 0.408 & 0.405 & 0.842 & 0.971 & 0.881 \\ 
  & EV*T3\_2 & 0.038 & 0.062 & 0.149 & 0.184 & 0.787 & 0.445 & 0.814 & 0.815 & 0.816 \\ 
  & H*T2\_1 & 0.228 & 0.467 & 0.458 & 0.674 & 0.288 & 0.310 & 0.709 & 0.595 & 0.734 \\ 
  & H*T3\_2 & 0.164 & 0.174 & 0.362 & 0.530 & 0.778 & 0.356 & 0.785 & 0.787 & 0.789 \\ 
  & EV*H*T2\_1 & 0.085 & 0.034 & 0.030 & 0.652 & 0.683 & 0.706 & 0.834 & 0.712 & 0.858 \\ 
  & EV*H*T3\_2 & 0.004 & 0.006 & 0.001 & 0.022 & 0.355 & 0.848 & 0.477 & 0.481 & 0.486 \\ 
  \hline
\end{tabular}}
\caption{Estimates and the corresponding p-values obtained using \texttt{lmer} for increasing $m$.}
\label{tab:cont-lmer-est}
\end{table}

\begin{table}[htbp]
  \centering
  \resizebox{\textwidth}{!}{
\begin{tabular}{rrrrrrrrrrr}
  \hline
 & & $m=1$ & $m=2$ & $m=3$ & $m=4$ & $m=5$ & $m=6$ & $m=7$ & $m=8$ & $m=9$ \\ 
  \hline
$\beta$ & Intercept & 7.041 & 7.048 & 7.048 & 7.048 & 7.078 & 7.078 & 7.073 & 7.073 & 7.064 \\ 
  & EV & 7.884 & 7.900 & 7.900 & 7.900 & 7.890 & 7.894 & 7.900 & 7.900 & 7.888 \\ 
  & H & -0.001 & 0.000 & 0.000 & 0.000 & -0.005 & -0.003 & 0.008 & 0.008 & -0.006 \\ 
  & T2\_1 & -0.016 & -0.017 & -0.017 & -0.017 & -0.014 & -0.006 & 0.005 & 0.005 & 0.013 \\ 
  & T3\_2 & -0.033 & -0.042 & -0.042 & -0.042 & -0.047 & -0.056 & -0.064 & -0.064 & -0.055 \\ 
  & EV*H & 0.495 & 0.504 & 0.504 & 0.504 & 0.495 & 0.512 & 0.527 & 0.527 & 0.534 \\ 
  & EV*T2\_1 & 0.046 & 0.033 & 0.033 & 0.033 & 0.051 & 0.070 & 0.066 & 0.066 & 0.068 \\ 
  & EV*T3\_2 & -0.114 & -0.139 & -0.139 & -0.139 & -0.158 & -0.176 & -0.171 & -0.171 & -0.159 \\ 
  & H*T2\_1 & 0.099 & 0.078 & 0.078 & 0.078 & 0.104 & 0.128 & 0.129 & 0.129 & 0.132 \\ 
  & H*T3\_2 & -0.091 & -0.077 & -0.077 & -0.077 & -0.108 & -0.126 & -0.117 & -0.117 & -0.104 \\ 
  & EV*H*T2\_1 & 0.153 & 0.096 & 0.096 & 0.096 & 0.139 & 0.170 & 0.174 & 0.174 & 0.223 \\ 
  & EV*H*T3\_2 & -0.319 & -0.277 & -0.277 & -0.277 & -0.312 & -0.352 & -0.338 & -0.338 & -0.360 \\
  \hline
  $\sigma^2$ & Intercept & 0.130 & 0.138 & 0.138 & 0.138 & 0.111 & 0.120 & 0.129 & 0.129 & 0.138 \\ 
  & EV & 0.129 & 0.128 & 0.128 & 0.128 & 0.133 & 0.142 & 0.151 & 0.151 & 0.178 \\ 
  & H & 0.071 & 0.078 & 0.078 & 0.078 & 0.083 & 0.092 & 0.096 & 0.096 & 0.088 \\ 
  & T2\_1 & 0.010 & 0.010 & 0.010 & 0.010 & 0.014 & 0.016 & 0.016 & 0.016 & 0.016 \\ 
  & T3\_2 & 0.005 & 0.005 & 0.005 & 0.005 & 0.009 & 0.012 & 0.017 & 0.017 & 0.028 \\ 
  & EV*H & 0.063 & 0.071 & 0.071 & 0.071 & 0.082 & 0.085 & 0.087 & 0.087 & 0.090 \\ 
  & EV*T2\_1 & 0.000 & 0.000 & 0.000 & 0.000 & 0.000 & 0.000 & 0.000 & 0.000 & 0.000 \\ 
  & EV*T3\_2 & 0.101 & 0.070 & 0.070 & 0.070 & 0.080 & 0.089 & 0.104 & 0.104 & 0.134 \\ 
  & H*T2\_1 & 0.000 & 0.000 & 0.000 & 0.000 & 0.000 & 0.000 & 0.000 & 0.000 & 0.000 \\ 
  & H*T3\_2 & 0.000 & 0.000 & 0.000 & 0.000 & 0.006 & 0.021 & 0.022 & 0.022 & 0.018 \\ 
  & $\eta_0$ & 0.121 & 0.125 & 0.125 & 0.125 & 0.126 & 0.129 & 0.138 & 0.138 & 0.146 \\
  \hline
  $p$-value & Intercept.1 & 0.000 & 0.000 & 0.000 & 0.000 & 0.000 & 0.000 & 0.000 & 0.000 & 0.000 \\ 
  & EV & 0.000 & 0.000 & 0.000 & 0.000 & 0.000 & 0.000 & 0.000 & 0.622 & 0.000 \\ 
  & H & 0.989 & 0.998 & 1.000 & 0.999 & 0.971 & 0.976 & 0.948 & 0.984 & 0.978 \\ 
  & T2\_1 & 0.784 & 0.842 & 0.849 & 0.835 & 0.920 & 0.971 & 0.975 & 0.973 & 0.932 \\ 
  & T3\_2 & 0.600 & 0.771 & 0.681 & 0.745 & 0.752 & 0.764 & 0.726 & 0.841 & 0.763 \\ 
  & EV*H & 0.000 & 0.000 & 0.001 & 0.001 & 0.010 & 0.018 & 0.018 & 0.128 & 0.040 \\ 
  & EV*T2\_1 & 0.672 & 0.802 & 0.847 & 0.878 & 0.903 & 0.820 & 0.825 & 0.931 & 0.827 \\ 
  & EV*T3\_2 & 0.348 & 0.496 & 0.363 & 0.435 & 0.758 & 0.708 & 0.612 & 0.907 & 0.831 \\ 
  & H*T2\_1 & 0.178 & 0.798 & 0.669 & 0.692 & 0.694 & 0.646 & 0.679 & 0.909 & 0.686 \\ 
  & H*T3\_2 & 0.263 & 0.791 & 0.898 & 0.725 & 0.639 & 0.581 & 0.653 & 0.906 & 0.694 \\ 
  & EV*H*T2\_1 & 0.359 & 0.894 & 0.821 & 0.820 & 0.782 & 0.766 & 0.775 & 0.812 & 0.724 \\ 
  & EV*H*T3\_2 & 0.069 & 0.712 & 0.472 & 0.527 & 0.492 & 0.497 & 0.677 & 0.929 & 0.669 \\ 
   \hline
\end{tabular}}
\caption{Estimates and the corresponding p-values obtained using the CVFS-estimator for increasing $m$.}
\label{tab:cont-cvfs-est}
\end{table}

\begin{table}[htbp]
  \centering
  \resizebox{\textwidth}{!}{
\begin{tabular}{rrrrrrrrrrr}
  \hline
 & & $m=1$ & $m=2$ & $m=3$ & $m=4$ & $m=5$ & $m=6$ & $m=7$ & $m=8$ & $m=9$ \\ 
  \hline
$\beta$ & Intercept & 7.062 & 7.062 & 7.055 & 7.040 & 7.059 & 7.060 & 7.082 & 7.078 & 7.070 \\ 
& EV & 7.836 & 7.836 & 7.835 & 7.843 & 7.837 & 7.838 & 7.828 & 7.841 & 7.856 \\ 
 & H & 0.002 & 0.003 & -0.003 & 0.012 & 0.006 & 0.007 & 0.004 & 0.018 & 0.026 \\ 
 & T2\_1 & -0.021 & -0.022 & -0.026 & -0.013 & -0.017 & -0.017 & -0.018 & -0.010 & 0.007 \\ 
 & T3\_2 & -0.030 & -0.028 & -0.019 & -0.023 & -0.027 & -0.028 & -0.029 & -0.033 & -0.042 \\ 
 & EV*H & 0.448 & 0.450 & 0.450 & 0.454 & 0.421 & 0.420 & 0.410 & 0.430 & 0.447 \\ 
 & EV*T2\_1 & 0.051 & 0.048 & 0.033 & 0.032 & 0.026 & 0.025 & 0.034 & 0.048 & 0.030 \\ 
 & EV*T3\_2 & -0.108 & -0.104 & -0.091 & -0.088 & -0.096 & -0.095 & -0.107 & -0.109 & -0.106 \\ 
 & H*T2\_1 & 0.116 & 0.115 & 0.113 & 0.122 & 0.120 & 0.120 & 0.140 & 0.148 & 0.158 \\ 
 & H*T3\_2 & -0.105 & -0.105 & -0.085 & -0.097 & -0.081 & -0.081 & -0.101 & -0.117 & -0.103 \\ 
 & EV*H*T2\_1 & 0.194 & 0.193 & 0.167 & 0.210 & 0.210 & 0.208 & 0.238 & 0.260 & 0.245 \\ 
 & EV*H*T3\_2 & -0.329 & -0.326 & -0.326 & -0.367 & -0.347 & -0.345 & -0.371 & -0.418 & -0.401 \\
  \hline
$\sigma^2$ & Intercept & 0.107 & 0.107 & 0.106 & 0.104 & 0.095 & 0.095 & 0.074 & 0.077 & 0.081 \\ 
  & EV & 0.137 & 0.137 & 0.142 & 0.146 & 0.151 & 0.150 & 0.150 & 0.151 & 0.154 \\ 
  & H & 0.064 & 0.064 & 0.065 & 0.061 & 0.063 & 0.062 & 0.064 & 0.061 & 0.063 \\ 
  & T2\_1 & 0.011 & 0.011 & 0.014 & 0.012 & 0.012 & 0.013 & 0.015 & 0.016 & 0.017 \\ 
  & T3\_2 & 0.016 & 0.015 & 0.016 & 0.018 & 0.018 & 0.018 & 0.021 & 0.022 & 0.023 \\ 
  & EV*H & 0.051 & 0.049 & 0.060 & 0.066 & 0.034 & 0.035 & 0.038 & 0.034 & 0.036 \\ 
  & EV*T2\_1 & 0.000 & 0.000 & 0.000 & 0.000 & 0.000 & 0.000 & 0.000 & 0.000 & 0.000 \\ 
  & EV*T3\_2 & 0.094 & 0.089 & 0.102 & 0.112 & 0.113 & 0.114 & 0.121 & 0.131 & 0.138 \\ 
  & H*T2\_1 & 0.000 & 0.000 & 0.000 & 0.000 & 0.000 & 0.000 & 0.000 & 0.000 & 0.001 \\ 
  & H*T3\_2 & 0.000 & 0.000 & 0.000 & 0.000 & 0.000 & 0.000 & 0.000 & 0.000 & 0.000 \\ 
  & $\eta_0$ & 0.109 & 0.109 & 0.104 & 0.105 & 0.107 & 0.106 & 0.103 & 0.106 & 0.113 \\ 
   \hline
$p$-value & Intercept & 0.000 & 0.000 & 0.000 & 0.000 & 0.000 & 0.000 & 0.000 & 0.000 & 0.000 \\ 
 & EV & 0.000 & 0.000 & 0.000 & 0.000 & 0.000 & 0.000 & 0.000 & 0.000 & 0.000 \\ 
 & H & 0.968 & 0.961 & 0.952 & 0.840 & 0.912 & 0.907 & 0.951 & 0.756 & 0.674 \\ 
 & T2\_1 & 0.685 & 0.662 & 0.610 & 0.799 & 0.748 & 0.745 & 0.746 & 0.862 & 0.903 \\ 
 & T3\_2 & 0.543 & 0.566 & 0.689 & 0.635 & 0.590 & 0.587 & 0.578 & 0.534 & 0.438 \\ 
 & EV*H & 0.000 & 0.000 & 0.000 & 0.000 & 0.000 & 0.000 & 0.000 & 0.000 & 0.000 \\ 
 & EV*T2\_1 & 0.514 & 0.536 & 0.670 & 0.686 & 0.746 & 0.758 & 0.674 & 0.563 & 0.724 \\ 
 & EV*T3\_2 & 0.268 & 0.284 & 0.353 & 0.384 & 0.354 & 0.358 & 0.306 & 0.313 & 0.347 \\ 
 & H*T2\_1 & 0.148 & 0.151 & 0.166 & 0.142 & 0.163 & 0.163 & 0.095 & 0.083 & 0.075 \\ 
 & H*T3\_2 & 0.210 & 0.211 & 0.299 & 0.243 & 0.334 & 0.337 & 0.213 & 0.153 & 0.217 \\ 
 & EV*H*T2\_1 & 0.256 & 0.259 & 0.328 & 0.218 & 0.232 & 0.236 & 0.172 & 0.150 & 0.193 \\ 
 & EV*H*T3\_2 & 0.029 & 0.030 & 0.034 & 0.016 & 0.025 & 0.025 & 0.017 & 0.005 & 0.009 \\ 
\hline
\end{tabular}}
\caption{Estimates and the corresponding p-values  obtained using the proposed MDPDE with $\alpha = 1/13$ for increasing $m$.}
\label{tab:cont-dpd-est}
\end{table}

Tables \ref{tab:cont-lmer-est}, \ref{tab:cont-cvfs-est} and \ref{tab:cont-dpd-est} display the estimates and the corresponding $p$-values obtained using
the usual \texttt{lmer}, the CVFS-estimator and the proposed MDPDE, respectively, for
increasing value of the number $(m)$ of substituted cells.

\bibliography{dpd-lmm}